\documentclass[twocolumn,prl,aps,showpacs,superscriptaddress,showkeys,reprint]{revtex4-1}
\usepackage{amssymb, amsmath}
\usepackage[english]{babel}
\usepackage{bm}
\usepackage{graphicx}
\usepackage[utf8]{inputenc}
\usepackage{color}

\newcommand\mydots{...\,}

\newcommand{\REVchange}[1]{#1} 

\begin{document}
%TC:ignore
\author{Sandra C. Kuhn}
\author{Marten Richter}
\email[]{marten.richter@tu-berlin.de}

\affiliation{Institut für Theoretische Physik, Nichtlineare Optik und
Quantenelektronik, Technische Universität Berlin, Hardenbergstr. 36, EW 7-1, 10623
Berlin, Germany}

\title{Combined tensor network/cluster expansion method using logic gates:\\ Illustrated for (bi-)excitons by a  single layer MoS$_2$ model system}

% Make the title.

% Place your abstract within the special {sciabstract} environment.
%TC:ignore
\begin{abstract}
Carriers such as electrons and holes inside the Brillouin zone of complex semiconducting materials can form bound states (excitons, biexcitons etc.). 
For obtaining the corresponding eigenstates (e.g. through Wannier or Bethe Salpeter equation) and dynamics (e.g. cluster expansion) the number of involved electrons and holes as well as the accuracy is limited by the appearing high dimensional tensors (i.e. wavefunctions or correlations).
These tensors can be efficiently represented and manipulated via tensor network methods.
We show how tensor networks formulated via classic logic gates can be used to treat electron-hole complexes inside the Brillouin zone.
The method is illustrated for the exciton and biexciton states of a single layer transition metal dichalcogenide MoS$_2$ like model system.   

%(Is: ???  Should:??? words max. ) 
 \end{abstract}
 %TC:endignore

\date{\today}
%TC:ignore
\maketitle
 %TC:endignore
%{\bf   word/3000 words max}

\begin{figure}[tb] 	
  \includegraphics[width=8.5cm]{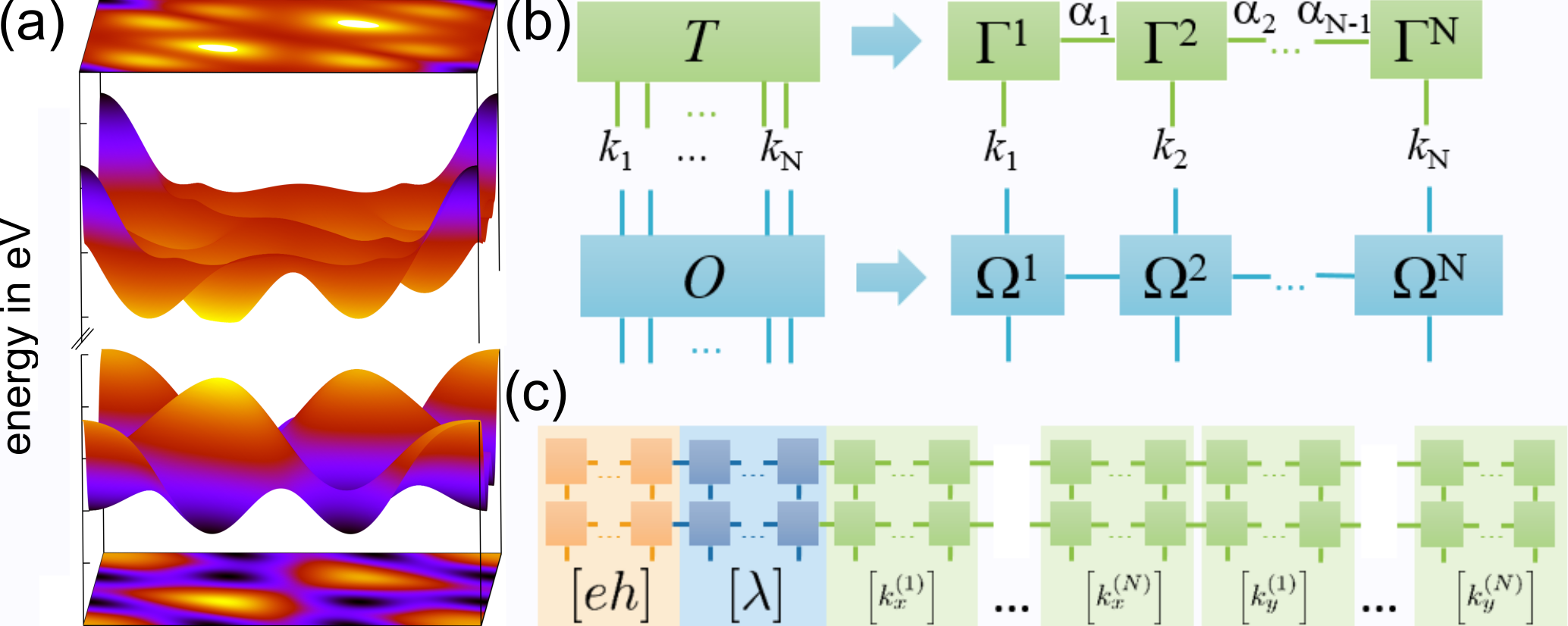}
  \caption{a) Valence and conduction band of MoS$_2$ for one spin. b) Decomposition of a tensor into a tensor train c) Diagrammatic representation of the tensor train coding  $S(\mathbf{k}_1^L\mydots \mathbf{k}_n^L| \mathbf{k}_1^R \mydots \mathbf{k}_m^R)$.}
  \label{tensorillu}
 \end{figure}

Semiconductor Bloch-equations and cluster expansion methods have been the work horse for optical induced 
carrier and exciton dynamics in semiconductor materials for decades \cite{lindberg1988effective,butscher2005ultrafast,reiter2007spatiotemporal,schmidt2016ultrafast,PhysRevLett.92.067402,winzer2010carrier,PhysRevB.97.035425} . Even for the recent monolayer two dimensional materials like monolayer TMDCs, these methods are still successful.
However most treatments of Coulomb bound electron-hole states were restricted to exciton states.
Trion and biexciton states and beyond are seldom included \cite{mostaani2017diffusion,ostreich1998theory,schmidt2016ultrafast,zimmermanntrion2000,esser2001theory,mackdropleton,steinhoff2018biexciton}.  If they are included,  the correlations are expanded in a basis of few  bound exciton, trion or biexciton states \cite{mostaani2017diffusion,ostreich1998theory,schmidt2016ultrafast,zimmermanntrion2000,esser2001theory,steinhoff2018biexciton}.\\
Exciton states are calculated using the Wannier equation \cite{haug2009quantum,berghauser2014analytical,mostaani2017diffusion,schmidt2016ultrafast,richter2017nanoplatelets}, either in real space \cite{zimmermanntrion2000,richter2017nanoplatelets,PhysRevB.95.235307,mostaani2017diffusion} or in reciprocal space \cite{haug2009quantum,huser2013dielectric,berghauser2014analytical}. Often the evaluation is restricted for exciton living near high symmetrical points such as the $\Gamma$ or $K$ and $K'$ points. In the context of ab initio treatments the equivalent Bethe-Salpeter equation (BSE) is used for calculating exciton states \cite{huser2013dielectric,PhysRevLett.108.126404,PhysRevLett.104.226804,qiu2013optical,laskowski2005ab,qiu2015nonanalyticity,PhysRevB.97.245427}. On the other hand calculations for higher order correlated electron hole states like trions or biexcitons are sparse \cite{stebe1997excitonic,mayrock1999weak,zimmermanntrion2000,esser2001theory,mostaani2017diffusion,PhysRevB.93.041401,10.1021acs.nanolett.8b00840,Drueppel2017,steinhoff2018biexciton}. This paper will illustrate a route to make bigger electron-hole complexes accessible: dynamically or for obtaining bound eigenstates. \\
Higher order correlations (induced by e.g. Coulomb, electron-phonon or electron-photon interaction) for few level systems like quantum dots have been successfully addressed by inductive equation of motion methods \cite{PhysRevB.84.125324,PhysRevLett.104.156801,PhysRevB.89.085308}.
However these methods are restricted to systems with few discrete levels and few discrete photon or phonon modes and cannot be applied so far to correlations with many continuous quasi momentum indices.
The required  memory sizes for storing the correlations scales exponentially in the number of involved particles $N$ and polynomial in the number of involved states $M$ ($\propto M^N$). 
Even small numbers of involved particles e.g. $N=4$ lead for a treatment of two dimensional Brillouin zone (BZ) (e.g. for small $30$ momentum points in each dimension a two band model yields $M=1800$) leads to a hard numerical problem.
So the stored data required for the simulation is the bottleneck to attack higher order correlations. However the naive raw data amount required to store entire movies on our computers and smartphones is at least impractical, but lossy data compression solves this issue and storing hundreds of movies on a single computer is possible. So for treating higher order many particle correlations a lossy data compression method 
and the ability to calculate directly on the compressed data will be the solution. 
Expanding the correlations in a basis (e.g. exciton \cite{ostreich1998theory,schmidt2016ultrafast}, trion \cite{stebe1997excitonic,zimmermanntrion2000,esser2001theory}, biexciton \cite{mayrock1999weak,steinhoff2018biexciton}, permutational symmetric basis \cite{gegg2016,Gegg2017}) is in principle already a first simple form of datacompression, where known symmetries and properties of the problem are used for an efficient description of the system.
However for every problem a different or modified basis is required, where the reformulation and implementation of the equations is tedious and requires  substantial effort.
In the context of highly correlated quantum systems tensor network methods like matrix product states (MPS) 
provided a systematic and reliable way to store and manipulate quantum states of e.g. spin chains \cite{verstraete2006matrix,vidal2007classical,CIRAC2017100,plenioheisenberg}, and also system bath interactions \cite{PhysRevLett.116.237201,shi2018}.
The wavefunction of the spin chain is interpreted as a tensor and decomposed in a tensor network such as a MPS (also called a tensor train (TT)).
In mathematics and chemistry a new trend uses TT (or other tensor networks) to compress high dimensional tensors regardless, if the tensor represents an actual quantum mechanical wavefunction \cite{shi2018}.\\
Furthermore for solving partial differential equations in real space quantics tensor trains (QTT) were introduced \cite{oseledets2009approximation,oseledets2010approximation,khoromskij2011dlog,kazeev2012low,khoromskaia2015tensor,benner2017fast}.
QTT do not use the spatial coordinates as indices of the tensors, but their binary representation.
We transfer this concept to cluster expansion and Wannier equations, since correlations appearing there are also tensors.
We show that a binary representation of the BZ quasi momentum allows a straight forward expression 
of the material equations using tensor networks with binary logic gates.
We focus on the calculation for a model system with realistic numerical complexity, that describes excitons and biexcitons formed between valence and conduction band of the two dimensional TMDC MoS$_2$ \cite{hao2017neutral,berghauser2014analytical,zhang2015experimental,sie2015intervalley,PhysRevB.93.041401,olsen2016simple,PhysRevB.95.081301,steinhoff2018biexciton,maltedarkbright,mostaani2017diffusion}.  
\REVchange{Currently monolayer TMDC such as MoS$_2$ are intensively researched as new material for applications as (quantum) optical devices. In particular, very strong excitonic effects caused by the remarkably strong Coulomb interactions make monolayer TMDCs a very unique class of materials. The bandstructure of MoS$_2$ (cf. Fig. \ref{tensorillu}a)) shows several valence and conduction band extrema, which are relevant to the formation of bound electron-hole complexes (e.g. for the exciton ground state is formed by electrons and holes at the $K$ or $K'$ point). 
In particular, for larger electron-hole complexes  such as biexcitons or trions long range Coulomb interactions inside the full BZ become important.
A calculation of the full BZ has the advantage that all possible bound quasiparticles formed by Coulomb interaction are included, not only just the one localized at the usually investigated $K$ and $K'$ symmetry points.}\\
We demonstrate that the electrons and holes complex quantities can be calculated using a very high number of grid points in the BZ.
This paper is a proof of principle, that the combination of tensor network methods increases the range of problems addressable with Wannier equation and BSE. 
\REVchange{The MoS$_2$ is chosen as model system, since it is one of the most widely studied TMD and the tight binding bandstructure is available from the literature \cite{ridolfi2015tight,ridolfi2018exstruc} as solid basis for the model system.}
Beside the calculation of (bi-)exciton states the concept is also extensible towards solving equations of motions in cluster expansion.\\
\noindent\textit{Model System:} The Hamilton operator $H$ of the model system is $H=H_0+H_C$.
The electronic bandstructure enters the Hamiltonian through $H_0=\hbar\sum_{\mathbf{k}\lambda} \varepsilon_{\mathbf{k}}^\lambda 
a^\dagger_{\mathbf{k}\lambda} a_{\mathbf{k}\lambda}$, here $\mathbf{k}$ is the quasi momentum in BZ, and $\lambda$ describes band and spin. $\varepsilon_{\mathbf{k}}^\lambda$ is the bandstructure of the material, for this paper the tight binding  bandstructure for MoS$_2$ from \cite{ridolfi2015tight,ridolfi2018exstruc} is used. 
\REVchange{Depending on the band $\lambda=c_\sigma,v_\sigma$ distinguishing conduction $c$ and valence band $v$, $a^\dagger_{\mathbf{k}\lambda}$ with spin $\sigma=\uparrow,\downarrow$ , $a_{\mathbf{k}\lambda}$ are the creation and annihilation operator of an electron (conduction band) or hole (valence band): $a^\dagger_{\mathbf{k}\lambda_\sigma=c_\sigma}=e^\dagger_{\mathbf{k}\sigma}$ and  $a^\dagger_{\mathbf{k}\lambda_\sigma=v_\sigma}=h^\dagger_{\mathbf{k}\sigma}$.}
The Coulomb interaction Hamiltonian $H_C$ reads $H_C=\sum_{\mathbf{k}_1\mathbf{k}_2\mathbf{q}\lambda_1\lambda_{2}} I_{\lambda_1\lambda_2}^{\mathbf{k}_1\mathbf{k}_2\mathbf{q}} a^\dagger_{\mathbf{k}_1\lambda_1} a^\dagger_{\mathbf{k}_2\lambda_2}a_{\mathbf{k}_2+\mathbf{q}\lambda_2} a_{\mathbf{k}_1-C_{\lambda_1}^{\lambda_2}\mathbf{q}\lambda_1}$. \\
$C_{\lambda_1}^{\lambda_2}$ is $1$, if $\lambda_1$ and $\lambda_2$ are both holes or both electron and $-1$ otherwise. 
The prefactor $I_{\lambda_1\lambda_2}^{\mathbf{k}_1\mathbf{k}_2\mathbf{q}}=F_{\mathbf{q}}^{\mathbf{k}_1\mathbf{k}_2} V_{\mathbf{q}}C_{\lambda_1}^{\lambda_2}$ includes the Keldysh style Coulomb potential $V_{\mathbf{q}}$ \cite{PhysRevB.88.045318,berghauser2014analytical} and the tight-binding (TB) coefficients $c_{\mathbf{k} n_1}$ inside $F_{\mathbf{q}}^{\mathbf{k}_1\mathbf{k}_2}=\sum_{n_1,n_2} c_{\mathbf{k}_1 n_1}^{*} c^{*}_{\mathbf{k}_2n_2} c_{\mathbf{k}_2+\mathbf{q} n_2} c_{\mathbf{k}_1-C_{\lambda_1}^{\lambda_2}\mathbf{q}n_1}$.
\REVchange{The Coulomb potential $V_{\mathbf{q}}$ is calculated for MoS$_2$ on silica substrate (air/silica interface).
The model system is slightly simplified (no exchange coupling term),
since this paper is focused on the method (more accurate treatments are subject to future studies).}\\

We introduce the multiindex $\mathbf{k}=\{\mathbf{k}\lambda\}$, $\lambda$ is only written explicitly, if needed.
The correlations describing the system are $\langle a^\dagger_{\mathbf{k}_1^L}\mydots a^\dagger_{\mathbf{k}_n^L} a_{\mathbf{k}_m^R}\dots a_{\mathbf{k}_1^R}\rangle=:S(\mathbf{k}_1^L\mydots\mathbf{k}_n^L| \mathbf{k}_1^R\mydots\mathbf{k}_m^R)$.
Using Heisenberg equations of motion $\partial_t\langle O\rangle=i/\hbar \langle [H,O]_-\rangle$, 
we arrive an equation for $S$:
\begin{eqnarray}
 &&\partial_t S(\mathbf{k}_1^L\mydots \mathbf{k}_n^L| \mathbf{k}_1^R \mydots \mathbf{k}_m^R)=  i\sum_{j=1}^n \varepsilon_{\mathbf{k}_j^L} S(\mydots|\mydots)\nonumber\\
 &&\quad -2i \sum_{j \mathbf{k}\mathbf{q}}
 I_{\lambda_j^R\lambda}^{\mathbf{k}_j^R\mathbf{k}\mathbf{q}}
 S(\mydots\mathbf{k}_n^L \{\mathbf{k}+\mathbf{q}\}|\mydots\{\mathbf{k}_{j}^R-C^\lambda_{\lambda_j^R}\mathbf{q}\}\mydots\mathbf{k}_m^R \mathbf{k})\nonumber\\
 &&\quad -2i\sum_{i<j\mathbf{q}}
 I_{\lambda_j^R\lambda^R_i}^{\mathbf{k}_j^R\mathbf{k}_i^R\mathbf{q}}
 S(\mydots|\mydots\{\mathbf{k}_{j}^R-C^\lambda_{\lambda_j^R}\mathbf{q}\}\mydots\{\mathbf{k}_{i}^R+\mathbf{q}\}\mydots)\nonumber\\
 &&\quad -\{L\leftrightarrow R, n  \leftrightarrow m\}. \label{maineqmo}
\end{eqnarray}
On the rhs only the changes in the indices of $S(\mydots|\mydots)$ compared to the lhs are denoted. 
Note,  some terms on rhs change the number of indices of $S(\mydots|\mydots)$ compared to the lhs.  Eq. (\ref{maineqmo}) creates the usual infinite hierarchy of correlations, \REVchange{which is usually driven by electron-light interaction terms, not included here.}
Examples of higher order correlations include important spectroscopic contributions such as $S(\mathbf{k}_1\mathbf{k}_2 \mathbf{k}_3| \mathbf{k}_4)$ (contains density assisted polarizations leading to excitation induced dephasing), $S(\mathbf{k}_1\mathbf{k}_2 \mathbf{k}_3 \mathbf{k}_4|)$ (contains biexcitonic coherences) or $S(\mathbf{k}_1\mathbf{k}_2 \mathbf{k}_3 \mathbf{k}_4|\mathbf{k}_5 \mathbf{k}_6)$ (contains single exciton to biexciton correlations).  
These higher order tensors  $S(\mydots|\mydots)$ impose the high numerical burden, that tensor network methods will lift.
While for the future, calculation of the quantum dynamics using Eq. (\ref{maineqmo}) using the tensor network approach are possible, we focus in this paper on the calculation of many particle eigenstates, i.e. eigenstates for excitons, biexcitons etc.. 
Like in the calculation of the Wannier equation \cite{haug2009quantum}, we take the homogenous part of Eq. (\ref{maineqmo}) and convert the equation to an eigenproblem with eigenenergies $E$ for the respective many particle complexes:
 \begin{eqnarray}
  &&i E \,S(\mathbf{k}_1^L\mydots \mathbf{k}_n^L| \mathbf{k}_1^R \mydots \mathbf{k}_m^R)=  i\sum_{j=1}^n \varepsilon_{\mathbf{k}_j^L}
 S(\mydots|\mydots)\nonumber\\
  &&\quad -2i\sum_{i<j\mathbf{q}}
  I_{\lambda_j^R\lambda^R_i}^{\mathbf{k}_j^R\mathbf{k}_i^R\mathbf{q}}
  S(\mydots|\mydots\{\mathbf{k}_{j}^R-C^\lambda_{\lambda_j^R}\mathbf{q}\}\mydots\{\mathbf{k}_{i}^R+\mathbf{q}\}\mydots)\nonumber\\
  &&\quad -\{L\leftrightarrow R, n  \leftrightarrow m\}. \label{bethesalpeterlike}
 \end{eqnarray}
 \REVchange{Then $S(|\mathbf{k}_1 \dots \mathbf{k}_N)$ Eq. (\ref{bethesalpeterlike}) has actually the same form as the hermitian conjugate of the Schrödinger equation of $\Psi_{\mathbf{k}_1 \dots   \mathbf{k}_N}=a^\dagger_{\mathbf{k}_1}\dots a^\dagger_{\mathbf{k}_N}|\Psi_0\rangle$, which describes bound electron-hole carriers created out of the neutral ground state $|\Psi_0\rangle$ (without prior doping or optical excitation) of the system.
Therefore, for} $S(|\mathbf{k}_1 \mathbf{k}_2)$ Eq. (\ref{bethesalpeterlike}) corresponds to the Wannier equation, \REVchange{ the eigenstate problem for} excitons (bound electron hole pairs) \cite{haug2009quantum} in reciprocal space and is equivalent to a BSE \cite{qiu2013optical,10.1021acs.nanolett.8b00840,schmidt2016ultrafast}.
\REVchange{ Remember, 
 $\langle a_{\mathbf{k}v_{\sigma_1}} a_{\mathbf{k}c_{\sigma_2}} \rangle$ is part of $S(|\mathbf{k}_1 \mathbf{k}_2)$  and is an electron hole coherence $\langle h_{\mathbf{k}\sigma_1} e_{\mathbf{k}\sigma_2} \rangle$.}
\REVchange{ Furthermore,} for $S(|\mathbf{k}_1 \mathbf{k}_2 \mathbf{k}_3 \mathbf{k}_4)$ Eq. (\ref{bethesalpeterlike}) is the generalization to biexcitons, i.e. bound complexes from two electrons and two holes. In this paper, we will focus on excitons and biexcitons.

\noindent\textit{Tensor Network Methods:} In principle, $S(\mathbf{k}_1^L\mydots\mathbf{k}_n^L| \mathbf{k}_1^R\mydots\mathbf{k}_m^R)$ is a tensor 
 with indices $\mathbf{k}_1^L$,\dots, $\mathbf{k}_n^L$, $\mathbf{k}_1^R$,\dots, $\mathbf{k}_m^R$ and rank $n+m$.
 If we assume $g=1000$ grid points for the BZ (which is probably too small), the memory requirement is $g^{n+m}=1000^{n+m}$, so that already for very small $m$ and $n$ the memory exceeds the feasible and possible range.
  \cite{vidal2003efficient} showed that every tensor $T_{k_1,\dots,k_N}$ can be approximated as MPS (in mathematics TT) in the form: 
  $T_{k_1,\dots,k_n}=\sum_{\alpha_1,..,\alpha_{n-1}} \Gamma^{1,k_1}_{\alpha_1}\Gamma^{2,k_2}_{\alpha_1\alpha_2}\Gamma^{3,k_3}_{\alpha_2\alpha_3}
   \cdots \Gamma^{n,k_{n}}_{\alpha_{n-1}}$.  The tensors $\Gamma^{n,k}_{\alpha\alpha'}$ have a maximum of $g\cdot \REVchange{D}^2$ elements, if \REVchange{$D$} is the maximum number of $\alpha_i$ (link dimension).
   If the relevant information of the tensor can be represented with small \REVchange{$D$}, 
   the overall memory size reduces from exponential scaling $g^{n+m}$ 
   to linear scaling $(n+m)\cdot g\cdot \REVchange{D}^2$ making higher dimensional tensors accessible \cite{vidal2003efficient,orus2014practical,schollwock2011density,CIRAC2017100}.
   In the following we will use a diagrammatic notation for tensors \cite{orus2014practical,schollwock2011density,CIRAC2017100}: 
   The tensor is represented by a rectangle and indices are denoted as lines (cf. Fig. \ref{tensorillu} b)). 
   If the two indices of a tensor are contracted (summed) the lines are connected, so that the decomposition of the tensor $T_{k_1,\dots,k_n}$ into $\Gamma^{n,k}_{\alpha\alpha'}$ are represented by the diagram Fig. \ref{tensorillu} b).
   Tensors are mathematically vectors, simple vector operations such as adding, taking the norm, scalar multiplication can be carried out directly  on the MPS form without reconstructing the full tensor \cite{orus2014practical,schollwock2011density}. 
   Linear operators $O_{k_1,\dots,k_n,k_1',\dots,k_n'}$ acting on tensors represented as MPS  can be described as matrix product operators (MPO), which can be applied efficiently on MPS (see Fig. \ref{tensorillu} b) and \cite{orus2014practical,schollwock2011density,vidal2004efficient}).\\
  Representing $S(\mathbf{k}_1^L\mydots\mathbf{k}_n^L| \mathbf{k}_1^R\mydots\mathbf{k}_m^R)$ directly as MPS is not a good idea, since the dimension $g$ (the number of grid points) of $\mathbf{k}$ is very high.
  In \cite{oseledets2009approximation,oseledets2010approximation,khoromskij2011dlog,kazeev2012low,khoromskaia2015tensor} QTT were introduced to solve this problem, also in the context of BSEs \cite{benner2017fast}. For a QTT the tensor indices are not used for the decomposition, but the bits of a binary representation of the indices resulting in a $(n+m)\cdot \log(g)\cdot \REVchange{D}^2$ scaling of the memory requirement.
For a binary representation of the 2D BZ, the quasi momentum is written as $\mathbf{k}=1/{2^N} \sum_{i=1}^N ( k_x^{(i)} \mathbf{b}_x + k_y^{(i)} \mathbf{b}_y)  2^i$ with the \REVchange{number of bits $N$ and} the basis vectors $\mathbf{b}_{x/y}$ of the BZ and bits $k^{(i)}_{x/y}=0,1$.
Furthermore the band index $\lambda$ contains one bit $eh$ for distinguishing valence and conduction band and one bit for the spin $s$. The bit representation is very suitable for the interaction terms, since relations like quasi momentum conservation including Umklapp processes can be represented by binary logic gates inside tensor networks. 
Most binary logic operations between two $\mathbf{k}$'s connects bits from the same digit, or the adjacent digits.
Sorting the bit indices for the QTT/MPS decomposition by binary digits results in more efficient tensor networks. Therefore, the bits to describe the indices $\mathbf{k}_i^{L/R}$ of the tensor $S(\mathbf{k}_1^L \mydots \mathbf{k}_n^L| \mathbf{k}_1^R \mydots \mathbf{k}_m^R)$ are sorted as
($[eh]$, $[\lambda]$, $[k^{(1)}_x]$, \dots, $[k^{(N)}_x]$, $[k^{(1)}_y]$, \dots, $[k^{(N)}_y]$), where $[\,\cdot\,]$ represents a group of bits: $[\Lambda]=  \Lambda_1^L,\dots,\Lambda_n^L, \Lambda_m^R,\dots \Lambda_1^R$, cf. Fig. \ref{tensorillu} c). 
After defining the QTT decomposition, the MPOs are build from tensor networks for the rhs terms of Eq. (\ref{bethesalpeterlike}).

 \begin{figure}[tb] 	
  \includegraphics[width=8.5cm]{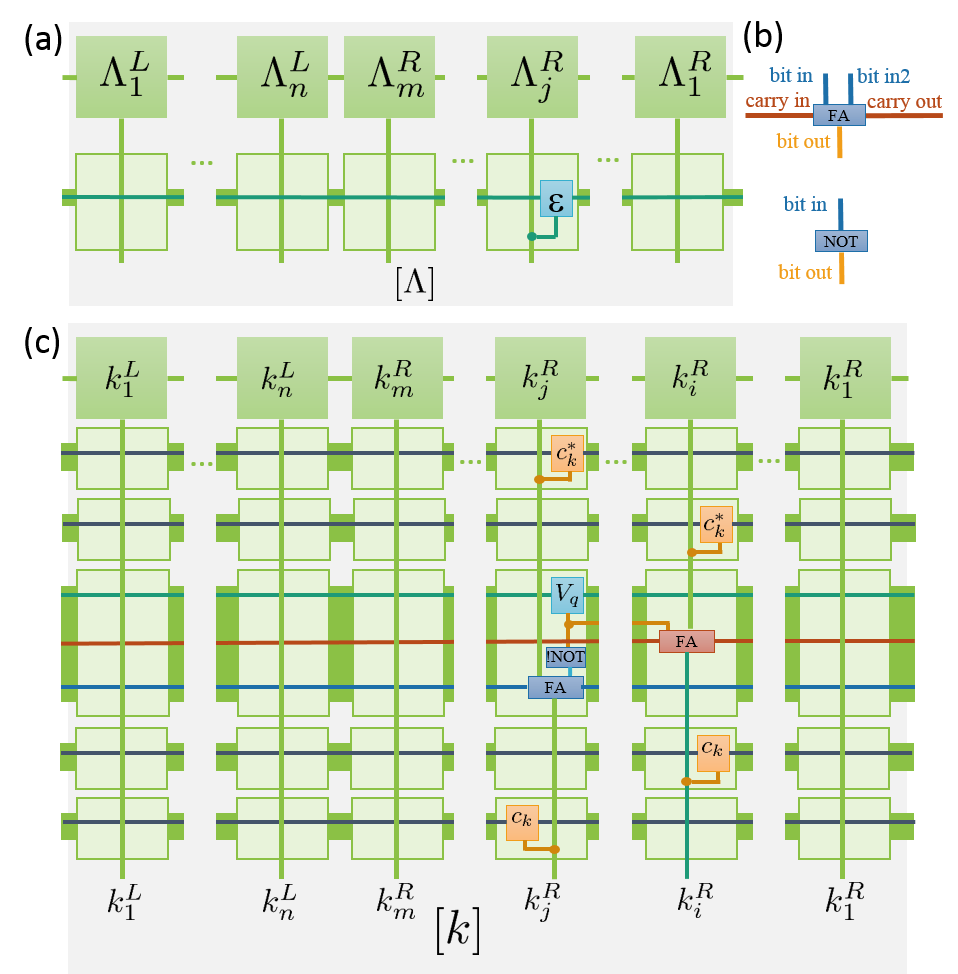}
  \caption{a) TN representing the homogeneous part of Eq. (\ref{bethesalpeterlike}),
  b) full adder (FA) and negation (NOT) logical circuits, c) TN representing the Coulomb term of Eq. (\ref{bethesalpeterlike}) (only $k$ part). }
  \label{TNeqmo}
 \end{figure} 

\noindent We start with the homogeneous energy term $-i\varepsilon_{\mathbf{k}^R_j}S(\mydots|\mydots\mathbf{k}^R_j\mydots)$, where $S(\mydots|\mydots\mathbf{k}^R_j\mydots)$ is the MPS, on which a MPO will act, see Fig. \ref{tensorillu} c).  
We can rewrite the term as $-i\sum_{\mathbf{k'}^R_j\mathbf{k''}^R_j}\delta_{\mathbf{k}^R_j\mathbf{k'}^R_j\mathbf{k''}^R_j} \,\varepsilon_{\mathbf{k'}^R_j}S(\mydots|\mydots\mathbf{k''}^R_j\mydots)$, where $\delta_{\mathbf{k}^R_j\mathbf{k'}^R_j\mathbf{k''}^R_j}$ factorizes into $\prod_i\delta_{k^{(i)R}_{mj}{k'}^{(i)R}_{mj}{k''}^{(i)R}_{mj}}$ for every bit of $\mathbf{k}^R_j$ including band and spin bits. Furthermore, $\varepsilon_{\mathbf{k}^R_j}$ is converted into a QTT ${\bf \varepsilon}$ with the same bit ordering. The tensor network in Fig. \ref{TNeqmo}a) depicts the MPO of the homogeneous energy term.
The $\delta-$ bit tensor is represented by a dot in Fig. \ref{TNeqmo} and connects the initial MPS  bit index with the  bit index of $\varepsilon_{\mathbf{k}^R_j}$ and the final bit index.
For every tensor network construction, the key design principle is to  ensure  the correct flow of index information from the initial MPS  ($S(...|...)$ on the rhs) to the term prefactors inside the MPO to the final MPS indices ($S(...|...)$ on the lhs).
 In the TN in Fig. \ref{TNeqmo}a) the connections and junction ensure that the same indices of the initial tensor on the rhs of Eq. (\ref{bethesalpeterlike}), the energy tensor $\varepsilon_{\mathbf{k}^R_j}$ and the tensor on the lhs are connected. \\
Constructing the TN for the Coulomb term $-2i\sum_{i<j\mathbf{q}}I_{\lambda_j^R\lambda^R_i}^{\mathbf{k}_j^R\mathbf{k}_i^R\mathbf{q}}\, S(\mydots|\mydots\{\mathbf{k}_{j}^R-C^\lambda_{\lambda_j^R}\mathbf{q}\}\mydots\{\mathbf{k}_{i}^R+\mathbf{q}\}\mydots)$ is more involved and will require five MPOs, which are subsequently compressed to a single MPO.
 $\mathbf{q}$ has positive and negative components, negative number are encoded using two's complement representation for binary negative integers \cite{tietze2015electronic}, which matches nicely the periodic properties of the BZ. (A negative $\mathbf{q}$ is represented by a positive $\mathbf{q}+\mathbf{G}$ inside BZ with suitable $\mathbf{G}$.)
 For the correct flow of information the TN has to connect the bit indices for $\mathbf{k}_{i}^R$, $\mathbf{k}_{j}^R$, $\mathbf{k}_{i}^R-\mathbf{q}$, $\mathbf{k}_{j}^R+C^\lambda_{\lambda_j^R}\mathbf{q}$ and $\mathbf{q}$ with $C^\lambda_{\lambda_j^R}=\pm 1$.
Tensors representing binary logic gates achieve this: a set of full adders \cite{tietze2015electronic}  calculate $\mathbf{k}_{i}^R-\mathbf{q}$ and $\mathbf{k}_{j}^R+C^\lambda_{\lambda_j^R}\mathbf{q}$ from $\mathbf{k}_{i}^R$,  $\mathbf{k}_{j}^R$ and $\mathbf{q}$. For the case $C^\lambda_{\lambda_j^R}=-1$ additional NOT circuits convert $\mathbf{q}$ to a negative input in two's complement representation for the full adder.
The corresponding TN is shown in Fig. \ref{TNeqmo}c), the application of $V_{\mathbf{q}}$ and calculation of the indices is handled by the MPO in the middle of the set of five MPOs.  Here the fulladders combine the $\mathbf{k}$ indices and the $\mathbf{q}$ indices of $V_{\mathbf{q}}$ for every bit of the binary representation. In addition carry bits connect the full adder for different bit digits.
In  Fig. \ref{TNeqmo}c) the prefactor $F^{\mathbf{k}_j^R\mathbf{k}_i^R}_{\mathbf{q}}$ is handled by the four outer MPO's.
For including it, the MPO from Fig. \ref{TNeqmo} is combined with two MPO's representing $c_{\mathbf{k}_2+\mathbf{q} n_2}$, $c_{\mathbf{k}_1-C_{\lambda_1}^{\lambda_2}\mathbf{q}n_1}$ before its application and two MPOs representing $c_{\mathbf{k}_1 n_1}^{*}$, $c^{*}_{\mathbf{k}_2n_2}$ after its application. (Supplemental material includes a more extensive discussion).\\
We use the ITensor C++ library (patched version 2.1.0),  for the calculation of all tensor operations \cite{itensor}.
The rhs of Eq. (\ref{bethesalpeterlike}) is calculated through the TN brought \REVchange{in the form of successive applied MPOs, which are compressed using a fitApply algorithm.}
In order to solve Eq. (\ref{bethesalpeterlike}) and to determine the respective exciton and biexciton eigenenergies and wavefunctions,\REVchange{ we first use a density matrix renormalization group (DMRG) algorithm} \cite{white1992density,schollwock2011density} which is capable of obtaining the eigenvalues and eigenvectors (MPS) of a MPO.  
We use a modified DMRG algorithm based on the itensor DMRG \cite{itensor} implementation for \REVchange{adding} multiple MPO and for calculation of higher energy eigenvectors.
\REVchange{ For the DMRG algorithm the successive applied set of MPOs (cf. Fig. \ref{TNeqmo} ) have to be merged into a single MPO, however the resulting MPO requires a very high link dimension and we could not achieve converged results. 
Using imaginary time propagation \cite{orus2014practical} for the final propagation  resulted in converged results, since here a merge of the subsequent applied MPOs is not necessary. (See convergence analysis in supplemental material.)
}
\REVchange{ For obtaining the eigenstates} Eq. (\ref{bethesalpeterlike}) is solved for the exciton $S(\mathbf{k}^L_1 \mathbf{k}^L_2|)$ and biexciton $S(\mathbf{k}^L_1 \mathbf{k}^L_2 \mathbf{k}^L_3 \mathbf{k}^L_4 | )$ coherences/wavefunction on a full $1024\times 1024$ grid for every $\mathbf{k}$-vector inside the full BZ.
To address optical excitability states, we focus on (bi-)exciton states with zero overall momentum \REVchange{$\langle a^\dagger_{\mathbf{k}v_{\sigma_1}} a^\dagger_{\mathbf{k}c_{\sigma_2}} \rangle 
= \langle h^\dagger_{\mathbf{k}\sigma_1} e^\dagger_{\mathbf{k}\sigma_2} \rangle$}
and \REVchange{$\langle a^\dagger_{\mathbf{k}v_{\sigma_1}} a^\dagger_{\mathbf{k}+\mathbf{q}c_{\sigma_2}} a^\dagger_{\mathbf{k}'+\mathbf{q}v_{\sigma_3}} a^\dagger_{\mathbf{k}'c_{\sigma_4}} \rangle 
= \langle h^\dagger_{\mathbf{k}\sigma_1} e^\dagger_{\mathbf{k}+\mathbf{q}\sigma_2} h^\dagger_{\mathbf{k}'+\mathbf{q}\sigma_3} e^\dagger_{\mathbf{k}'\sigma_4} \rangle$}
, the TN constructing this coherence from $S(\mathbf{k}^L_1 \mathbf{k}^L_2 \mathbf{k}^L_3 \mathbf{k}^L_4 | )$ is given in the Supplemental material.\\
%We obtain the bound, bright (dark) A and B exciton at \REVchange{$1.769~\mathrm{eV}$ ($1.805~\mathrm{eV}$) and $1.926~\mathrm{eV}$ ($1.924~\mathrm{eV}$)} 
We obtain the bound, bright  A exciton at1 \REVchange{$1.769~\mathrm{eV}$ }
compared to a band gap of $2.124~\mathrm{eV}$ at the K-point, reproducing \cite{ridolfi2018exstruc}, whose band structure \cite{ridolfi2015tight,ridolfi2018exstruc} is used in the model system, here.
\begin{figure}[tb] 	
  \includegraphics[width=8.5cm]{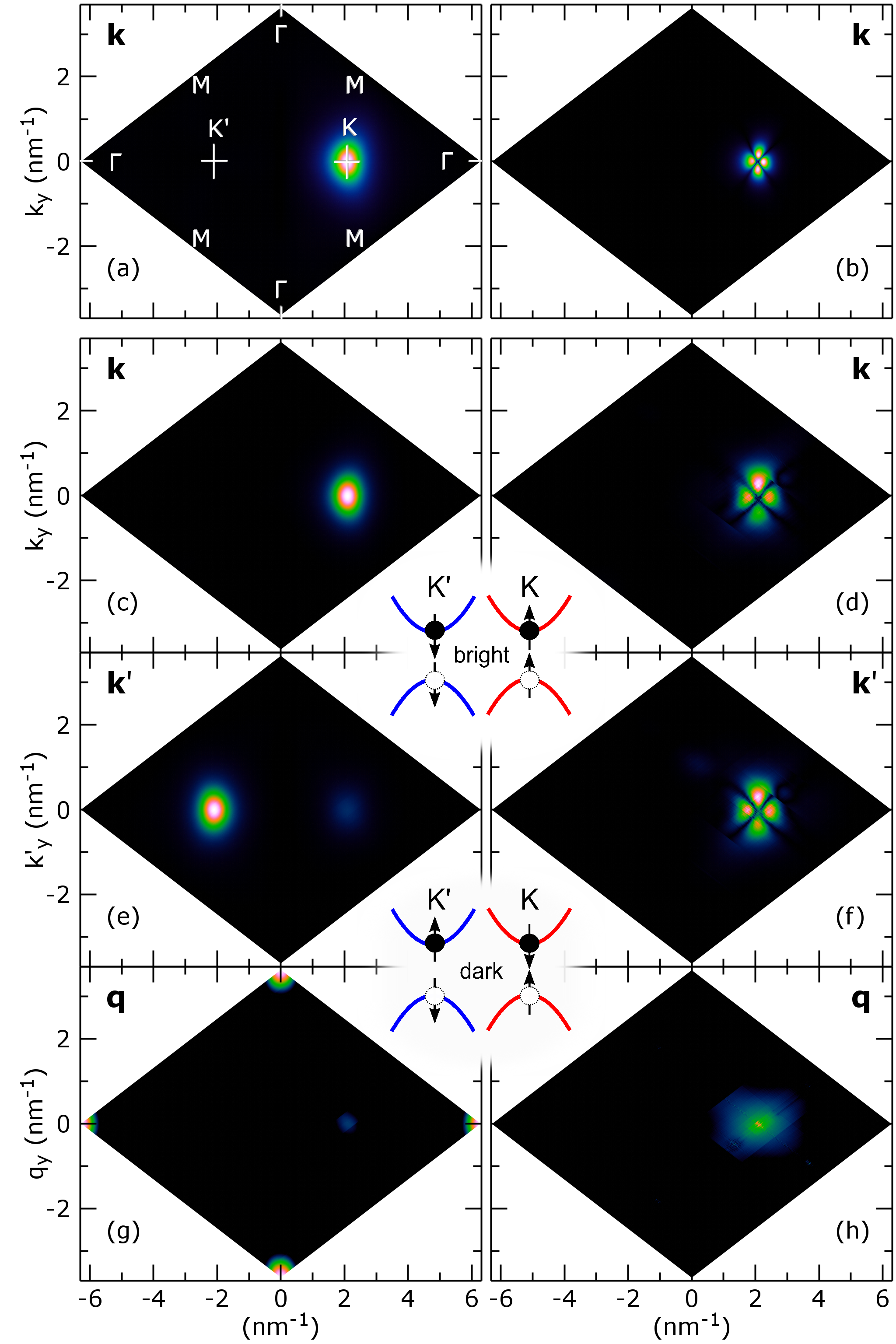}
  \caption{Selected exciton and biexciton wavefunctions in BZ: (a) bright A exciton with parallel spin up and (b) dark 3d exciton with anti parallel spin localized at the $K$-valley. (c), (e) and (g) bright biexciton formed from two bright A excitons and (d), (f) and (h) dark biexciton formed from two dark 3d excitons. The plotted variable is depicted in the corner. (g) and (h) are plotted using a logarithmic color scale. \REVchange{(a), (c), (e), (g) are converged results using imaginary time propagation and while (b), (d), (f), (h) use DMRG and show the potential of the method without achieving convergence.}}
  \label{results}
\end{figure} 
As an example from the exciton states, Fig. \ref{results}(a) shows the A exciton wavefunction \REVchange{($1.769~\mathrm{eV}$)} for parallel spin up configuration localized at the K-valley. 
The biexciton coherence with zero center of mass momentum \REVchange{$\langle a^\dagger_{\mathbf{k}v_{\sigma_1}} a^\dagger_{\mathbf{k}+\mathbf{q}c_{\sigma_2}} a^\dagger_{\mathbf{k}'+\mathbf{q}v_{\sigma_3}} a^\dagger_{\mathbf{k}'c_{\sigma_4}} \rangle$} depends on the three momenta $\mathbf{k}$, $\mathbf{k}^\prime$ and $\mathbf{q}$. To characterize the six dimensional wavefunction,  we sum over two momenta (e.g. $\mathbf{k}^\prime$ and $\mathbf{q}$) while plotting over the third (e.g. over $\mathbf{k}$) in the BZ. Fig. \ref{results}(c)-(h) shows two example biexciton states: 
Two A excitons, one electron-hole pair with parallel spin up located at the K- (Fig. \ref{results}(c)) and another with parallel spin down at the K'-valley (Fig. \ref{results}(e)), constitute a bright biexciton (note $\mathbf{q}\approx 0$, Fig. \ref{results}(g)) with an energy of \REVchange{$3.518~\mathrm{eV}$} (\REVchange{$20~\mathrm{meV}$} binding energy, cf. \cite{PhysRevB.95.081301,hao2017neutral}).\\
Furthermore, \REVchange{in principle} the approach allows also to access  higher energy bound biexciton states (bright or dark). 
Fig. \ref{results}(d),(f) and (h) shows as an example a biexciton composed from two dark 3d-excitons with anti-parallel electron-hole spin calculated using DMRG, (see wavefunction depicted in Fig. \ref{results}(b), cf. \cite{PhysRevB.93.235435}).
\REVchange{However using DMRG, the ordering of calculated higher excited states was highly parameter dependent and we could not achieve convergence, showing the need of modified DMRG types for excited states \cite{PhysRevB.94.045111,PhysRevLett.118.017201,PhysRevLett.116.247204} and successive applied MPOs.}
 Beside the example exciton and biexciton states, the framework allows to determine many higher energy bound electron holes states, in principle also for other correlated electron and hole quasi-particles like trions.\\
In conclusion, the combination of tensor networks, cluster expansion and logic gates on the Brillouin zone allows to easily access bound electron hole quasi-particle with little numerical cost and high precision.
\REVchange{We demonstrated our method on the example of excitonic and biexcitonic states in MoS$_2$ on a silicone substrate. The energies of the A and B excitons fit well with the results presented in Ref. \cite{ridolfi2018exstruc}, whose band structure is used in our model. The biexciton binding energy is in the same order of magnitude as reported in literature \cite{PhysRevB.95.081301,hao2017neutral}.}
Future studies in this framework will provide systematic investigation of the bound electron-hole complexes and extend the numerical technique to quantum dynamics.

\begin{acknowledgments}
\noindent We gratefully acknowledge support from the Deutsche Forschungsgemeinschaft (DFG) through SFB 787.
We thank E. Miles Stoudenmire for helpful tips and fast patches of the itensor library. We also acknowledge useful discussions with Jonathan Schwarz during his bachelor thesis in our group.
\end{acknowledgments}

%TC:ignore 
%COMMENT 500 words background COMMENT END
%TC:endignore 

%TC:endignore
%TC:ignore
%MAX: 30 References
%\bibliography{tensorbiblio}

\begin{thebibliography}{66}%
\makeatletter
\providecommand \@ifxundefined [1]{%
 \@ifx{#1\undefined}
}%
\providecommand \@ifnum [1]{%
 \ifnum #1\expandafter \@firstoftwo
 \else \expandafter \@secondoftwo
 \fi
}%
\providecommand \@ifx [1]{%
 \ifx #1\expandafter \@firstoftwo
 \else \expandafter \@secondoftwo
 \fi
}%
\providecommand \natexlab [1]{#1}%
\providecommand \enquote  [1]{``#1''}%
\providecommand \bibnamefont  [1]{#1}%
\providecommand \bibfnamefont [1]{#1}%
\providecommand \citenamefont [1]{#1}%
\providecommand \href@noop [0]{\@secondoftwo}%
\providecommand \href [0]{\begingroup \@sanitize@url \@href}%
\providecommand \@href[1]{\@@startlink{#1}\@@href}%
\providecommand \@@href[1]{\endgroup#1\@@endlink}%
\providecommand \@sanitize@url [0]{\catcode `\\12\catcode `\$12\catcode
  `\&12\catcode `\#12\catcode `\^12\catcode `\_12\catcode `\%12\relax}%
\providecommand \@@startlink[1]{}%
\providecommand \@@endlink[0]{}%
\providecommand \url  [0]{\begingroup\@sanitize@url \@url }%
\providecommand \@url [1]{\endgroup\@href {#1}{\urlprefix }}%
\providecommand \urlprefix  [0]{URL }%
\providecommand \Eprint [0]{\href }%
\providecommand \doibase [0]{http://dx.doi.org/}%
\providecommand \selectlanguage [0]{\@gobble}%
\providecommand \bibinfo  [0]{\@secondoftwo}%
\providecommand \bibfield  [0]{\@secondoftwo}%
\providecommand \translation [1]{[#1]}%
\providecommand \BibitemOpen [0]{}%
\providecommand \bibitemStop [0]{}%
\providecommand \bibitemNoStop [0]{.\EOS\space}%
\providecommand \EOS [0]{\spacefactor3000\relax}%
\providecommand \BibitemShut  [1]{\csname bibitem#1\endcsname}%
\let\auto@bib@innerbib\@empty
%</preamble>
\bibitem [{\citenamefont {Lindberg}\ and\ \citenamefont
  {Koch}(1988)}]{lindberg1988effective}%
  \BibitemOpen
  \bibfield  {author} {\bibinfo {author} {\bibfnamefont {M.}~\bibnamefont
  {Lindberg}}\ and\ \bibinfo {author} {\bibfnamefont {S.~W.}\ \bibnamefont
  {Koch}},\ }\href@noop {} {\bibfield  {journal} {\bibinfo  {journal} {Physical
  Review B}\ }\textbf {\bibinfo {volume} {38}},\ \bibinfo {pages} {3342}
  (\bibinfo {year} {1988})}\BibitemShut {NoStop}%
\bibitem [{\citenamefont {Butscher}\ \emph {et~al.}(2005)\citenamefont
  {Butscher}, \citenamefont {F{\"o}rstner}, \citenamefont {Waldm{\"u}ller},\
  and\ \citenamefont {Knorr}}]{butscher2005ultrafast}%
  \BibitemOpen
  \bibfield  {author} {\bibinfo {author} {\bibfnamefont {S.}~\bibnamefont
  {Butscher}}, \bibinfo {author} {\bibfnamefont {J.}~\bibnamefont
  {F{\"o}rstner}}, \bibinfo {author} {\bibfnamefont {I.}~\bibnamefont
  {Waldm{\"u}ller}}, \ and\ \bibinfo {author} {\bibfnamefont {A.}~\bibnamefont
  {Knorr}},\ }\href@noop {} {\bibfield  {journal} {\bibinfo  {journal}
  {Physical Review B}\ }\textbf {\bibinfo {volume} {72}},\ \bibinfo {pages}
  {045314} (\bibinfo {year} {2005})}\BibitemShut {NoStop}%
\bibitem [{\citenamefont {Reiter}\ \emph {et~al.}(2007)\citenamefont {Reiter},
  \citenamefont {Glanemann}, \citenamefont {Axt},\ and\ \citenamefont
  {Kuhn}}]{reiter2007spatiotemporal}%
  \BibitemOpen
  \bibfield  {author} {\bibinfo {author} {\bibfnamefont {D.}~\bibnamefont
  {Reiter}}, \bibinfo {author} {\bibfnamefont {M.}~\bibnamefont {Glanemann}},
  \bibinfo {author} {\bibfnamefont {V.}~\bibnamefont {Axt}}, \ and\ \bibinfo
  {author} {\bibfnamefont {T.}~\bibnamefont {Kuhn}},\ }\href@noop {} {\bibfield
   {journal} {\bibinfo  {journal} {Physical Review B}\ }\textbf {\bibinfo
  {volume} {75}},\ \bibinfo {pages} {205327} (\bibinfo {year}
  {2007})}\BibitemShut {NoStop}%
\bibitem [{\citenamefont {Schmidt}\ \emph {et~al.}(2016)\citenamefont
  {Schmidt}, \citenamefont {Bergh{\"a}user}, \citenamefont {Schneider},
  \citenamefont {Selig}, \citenamefont {Tonndorf}, \citenamefont {Malic},
  \citenamefont {Knorr}, \citenamefont {Michaelis~de Vasconcellos},\ and\
  \citenamefont {Bratschitsch}}]{schmidt2016ultrafast}%
  \BibitemOpen
  \bibfield  {author} {\bibinfo {author} {\bibfnamefont {R.}~\bibnamefont
  {Schmidt}}, \bibinfo {author} {\bibfnamefont {G.}~\bibnamefont
  {Bergh{\"a}user}}, \bibinfo {author} {\bibfnamefont {R.}~\bibnamefont
  {Schneider}}, \bibinfo {author} {\bibfnamefont {M.}~\bibnamefont {Selig}},
  \bibinfo {author} {\bibfnamefont {P.}~\bibnamefont {Tonndorf}}, \bibinfo
  {author} {\bibfnamefont {E.}~\bibnamefont {Malic}}, \bibinfo {author}
  {\bibfnamefont {A.}~\bibnamefont {Knorr}}, \bibinfo {author} {\bibfnamefont
  {S.}~\bibnamefont {Michaelis~de Vasconcellos}}, \ and\ \bibinfo {author}
  {\bibfnamefont {R.}~\bibnamefont {Bratschitsch}},\ }\href {\doibase
  10.1021/acs.nanolett.5b04733} {\bibfield  {journal} {\bibinfo  {journal}
  {Nano Letters}\ }\textbf {\bibinfo {volume} {16}},\ \bibinfo {pages} {2945}
  (\bibinfo {year} {2016})},\ \bibinfo {note} {pMID: 27086935},\ \Eprint
  {http://arxiv.org/abs/https://doi.org/10.1021/acs.nanolett.5b04733}
  {https://doi.org/10.1021/acs.nanolett.5b04733} \BibitemShut {NoStop}%
\bibitem [{\citenamefont {Chatterjee}\ \emph {et~al.}(2004)\citenamefont
  {Chatterjee}, \citenamefont {Ell}, \citenamefont {Mosor}, \citenamefont
  {Khitrova}, \citenamefont {Gibbs}, \citenamefont {Hoyer}, \citenamefont
  {Kira}, \citenamefont {Koch}, \citenamefont {Prineas},\ and\ \citenamefont
  {Stolz}}]{PhysRevLett.92.067402}%
  \BibitemOpen
  \bibfield  {author} {\bibinfo {author} {\bibfnamefont {S.}~\bibnamefont
  {Chatterjee}}, \bibinfo {author} {\bibfnamefont {C.}~\bibnamefont {Ell}},
  \bibinfo {author} {\bibfnamefont {S.}~\bibnamefont {Mosor}}, \bibinfo
  {author} {\bibfnamefont {G.}~\bibnamefont {Khitrova}}, \bibinfo {author}
  {\bibfnamefont {H.~M.}\ \bibnamefont {Gibbs}}, \bibinfo {author}
  {\bibfnamefont {W.}~\bibnamefont {Hoyer}}, \bibinfo {author} {\bibfnamefont
  {M.}~\bibnamefont {Kira}}, \bibinfo {author} {\bibfnamefont {S.~W.}\
  \bibnamefont {Koch}}, \bibinfo {author} {\bibfnamefont {J.~P.}\ \bibnamefont
  {Prineas}}, \ and\ \bibinfo {author} {\bibfnamefont {H.}~\bibnamefont
  {Stolz}},\ }\href {\doibase 10.1103/PhysRevLett.92.067402} {\bibfield
  {journal} {\bibinfo  {journal} {Phys. Rev. Lett.}\ }\textbf {\bibinfo
  {volume} {92}},\ \bibinfo {pages} {067402} (\bibinfo {year}
  {2004})}\BibitemShut {NoStop}%
\bibitem [{\citenamefont {Winzer}\ \emph {et~al.}(2010)\citenamefont {Winzer},
  \citenamefont {Knorr},\ and\ \citenamefont {Malic}}]{winzer2010carrier}%
  \BibitemOpen
  \bibfield  {author} {\bibinfo {author} {\bibfnamefont {T.}~\bibnamefont
  {Winzer}}, \bibinfo {author} {\bibfnamefont {A.}~\bibnamefont {Knorr}}, \
  and\ \bibinfo {author} {\bibfnamefont {E.}~\bibnamefont {Malic}},\
  }\href@noop {} {\bibfield  {journal} {\bibinfo  {journal} {Nano letters}\
  }\textbf {\bibinfo {volume} {10}},\ \bibinfo {pages} {4839} (\bibinfo {year}
  {2010})}\BibitemShut {NoStop}%
\bibitem [{\citenamefont {Meckbach}\ \emph {et~al.}(2018)\citenamefont
  {Meckbach}, \citenamefont {Stroucken},\ and\ \citenamefont
  {Koch}}]{PhysRevB.97.035425}%
  \BibitemOpen
  \bibfield  {author} {\bibinfo {author} {\bibfnamefont {L.}~\bibnamefont
  {Meckbach}}, \bibinfo {author} {\bibfnamefont {T.}~\bibnamefont {Stroucken}},
  \ and\ \bibinfo {author} {\bibfnamefont {S.~W.}\ \bibnamefont {Koch}},\
  }\href {\doibase 10.1103/PhysRevB.97.035425} {\bibfield  {journal} {\bibinfo
  {journal} {Phys. Rev. B}\ }\textbf {\bibinfo {volume} {97}},\ \bibinfo
  {pages} {035425} (\bibinfo {year} {2018})}\BibitemShut {NoStop}%
\bibitem [{\citenamefont {Mostaani}\ \emph {et~al.}(2017)\citenamefont
  {Mostaani}, \citenamefont {Szyniszewski}, \citenamefont {Price},
  \citenamefont {Maezono}, \citenamefont {Danovich}, \citenamefont {Hunt},
  \citenamefont {Drummond},\ and\ \citenamefont
  {Fal'Ko}}]{mostaani2017diffusion}%
  \BibitemOpen
  \bibfield  {author} {\bibinfo {author} {\bibfnamefont {E.}~\bibnamefont
  {Mostaani}}, \bibinfo {author} {\bibfnamefont {M.}~\bibnamefont
  {Szyniszewski}}, \bibinfo {author} {\bibfnamefont {C.}~\bibnamefont {Price}},
  \bibinfo {author} {\bibfnamefont {R.}~\bibnamefont {Maezono}}, \bibinfo
  {author} {\bibfnamefont {M.}~\bibnamefont {Danovich}}, \bibinfo {author}
  {\bibfnamefont {R.}~\bibnamefont {Hunt}}, \bibinfo {author} {\bibfnamefont
  {N.}~\bibnamefont {Drummond}}, \ and\ \bibinfo {author} {\bibfnamefont
  {V.}~\bibnamefont {Fal'Ko}},\ }\href@noop {} {\bibfield  {journal} {\bibinfo
  {journal} {Physical Review B}\ }\textbf {\bibinfo {volume} {96}},\ \bibinfo
  {pages} {075431} (\bibinfo {year} {2017})}\BibitemShut {NoStop}%
\bibitem [{\citenamefont {Östreich}\ \emph {et~al.}(1998)\citenamefont
  {Östreich}, \citenamefont {Schönhammer},\ and\ \citenamefont
  {Sham}}]{ostreich1998theory}%
  \BibitemOpen
  \bibfield  {author} {\bibinfo {author} {\bibfnamefont {T.}~\bibnamefont
  {Östreich}}, \bibinfo {author} {\bibfnamefont {K.}~\bibnamefont
  {Schönhammer}}, \ and\ \bibinfo {author} {\bibfnamefont {L.}~\bibnamefont
  {Sham}},\ }\href@noop {} {\bibfield  {journal} {\bibinfo  {journal} {Physical
  Review B}\ }\textbf {\bibinfo {volume} {58}},\ \bibinfo {pages} {12920}
  (\bibinfo {year} {1998})}\BibitemShut {NoStop}%
\bibitem [{\citenamefont {Esser}\ \emph {et~al.}(2000)\citenamefont {Esser},
  \citenamefont {Runge}, \citenamefont {Zimmermann},\ and\ \citenamefont
  {Langbein}}]{zimmermanntrion2000}%
  \BibitemOpen
  \bibfield  {author} {\bibinfo {author} {\bibfnamefont {A.}~\bibnamefont
  {Esser}}, \bibinfo {author} {\bibfnamefont {E.}~\bibnamefont {Runge}},
  \bibinfo {author} {\bibfnamefont {R.}~\bibnamefont {Zimmermann}}, \ and\
  \bibinfo {author} {\bibfnamefont {W.}~\bibnamefont {Langbein}},\ }\href
  {\doibase 10.1002/1521-396X(200003)178:1<489::AID-PSSA489>3.0.CO;2-R}
  {\bibfield  {journal} {\bibinfo  {journal} {physica status solidi (a)}\
  }\textbf {\bibinfo {volume} {178}},\ \bibinfo {pages} {489} (\bibinfo {year}
  {2000})}\BibitemShut {NoStop}%
\bibitem [{\citenamefont {Esser}\ \emph {et~al.}(2001)\citenamefont {Esser},
  \citenamefont {Zimmermann},\ and\ \citenamefont {Runge}}]{esser2001theory}%
  \BibitemOpen
  \bibfield  {author} {\bibinfo {author} {\bibfnamefont {A.}~\bibnamefont
  {Esser}}, \bibinfo {author} {\bibfnamefont {R.}~\bibnamefont {Zimmermann}}, \
  and\ \bibinfo {author} {\bibfnamefont {E.}~\bibnamefont {Runge}},\
  }\href@noop {} {\bibfield  {journal} {\bibinfo  {journal} {physica status
  solidi (b)}\ }\textbf {\bibinfo {volume} {227}},\ \bibinfo {pages} {317}
  (\bibinfo {year} {2001})}\BibitemShut {NoStop}%
\bibitem [{\citenamefont {Almand-Hunter}\ \emph {et~al.}(2014)\citenamefont
  {Almand-Hunter}, \citenamefont {Li}, \citenamefont {Cundiff}, \citenamefont
  {Mootz}, \citenamefont {Kira},\ and\ \citenamefont {Koch}}]{mackdropleton}%
  \BibitemOpen
  \bibfield  {author} {\bibinfo {author} {\bibfnamefont {A.~E.}\ \bibnamefont
  {Almand-Hunter}}, \bibinfo {author} {\bibfnamefont {H.}~\bibnamefont {Li}},
  \bibinfo {author} {\bibfnamefont {S.~T.}\ \bibnamefont {Cundiff}}, \bibinfo
  {author} {\bibfnamefont {M.}~\bibnamefont {Mootz}}, \bibinfo {author}
  {\bibfnamefont {M.}~\bibnamefont {Kira}}, \ and\ \bibinfo {author}
  {\bibfnamefont {S.~W.}\ \bibnamefont {Koch}},\ }\href {\doibase
  http://dx.doi.org/10.1038/nature12994 10.1038/nature12994} {\bibfield
  {journal} {\bibinfo  {journal} {Nature}\ }\textbf {\bibinfo {volume} {506}},\
  \bibinfo {pages} {471} (\bibinfo {year} {2014})}\BibitemShut {NoStop}%
\bibitem [{\citenamefont {Steinhoff}\ \emph {et~al.}(2018)\citenamefont
  {Steinhoff}, \citenamefont {Florian}, \citenamefont {Singh}, \citenamefont
  {Tran}, \citenamefont {Kolarczik}, \citenamefont {Helmrich}, \citenamefont
  {Achtstein}, \citenamefont {Woggon}, \citenamefont {Owschimikow},
  \citenamefont {Jahnke},\ and\ \citenamefont {Li}}]{steinhoff2018biexciton}%
  \BibitemOpen
  \bibfield  {author} {\bibinfo {author} {\bibfnamefont {A.}~\bibnamefont
  {Steinhoff}}, \bibinfo {author} {\bibfnamefont {M.}~\bibnamefont {Florian}},
  \bibinfo {author} {\bibfnamefont {A.}~\bibnamefont {Singh}}, \bibinfo
  {author} {\bibfnamefont {K.}~\bibnamefont {Tran}}, \bibinfo {author}
  {\bibfnamefont {M.}~\bibnamefont {Kolarczik}}, \bibinfo {author}
  {\bibfnamefont {S.}~\bibnamefont {Helmrich}}, \bibinfo {author}
  {\bibfnamefont {A.~W.}\ \bibnamefont {Achtstein}}, \bibinfo {author}
  {\bibfnamefont {U.}~\bibnamefont {Woggon}}, \bibinfo {author} {\bibfnamefont
  {N.}~\bibnamefont {Owschimikow}}, \bibinfo {author} {\bibfnamefont
  {F.}~\bibnamefont {Jahnke}}, \ and\ \bibinfo {author} {\bibfnamefont
  {X.}~\bibnamefont {Li}},\ }\href@noop {} {\bibfield  {journal} {\bibinfo
  {journal} {arXiv preprint arXiv:1801.04225}\ } (\bibinfo {year}
  {2018})}\BibitemShut {NoStop}%
\bibitem [{\citenamefont {Haug}\ and\ \citenamefont
  {Koch}(2009)}]{haug2009quantum}%
  \BibitemOpen
  \bibfield  {author} {\bibinfo {author} {\bibfnamefont {H.}~\bibnamefont
  {Haug}}\ and\ \bibinfo {author} {\bibfnamefont {S.~W.}\ \bibnamefont
  {Koch}},\ }\href@noop {} {\emph {\bibinfo {title} {Quantum Theory of the
  Optical and Electronic Properties of Semiconductors: Fivth Edition}}}\
  (\bibinfo  {publisher} {World Scientific Publishing Company},\ \bibinfo
  {year} {2009})\BibitemShut {NoStop}%
\bibitem [{\citenamefont {Bergh{\"a}user}\ and\ \citenamefont
  {Malic}(2014)}]{berghauser2014analytical}%
  \BibitemOpen
  \bibfield  {author} {\bibinfo {author} {\bibfnamefont {G.}~\bibnamefont
  {Bergh{\"a}user}}\ and\ \bibinfo {author} {\bibfnamefont {E.}~\bibnamefont
  {Malic}},\ }\href@noop {} {\bibfield  {journal} {\bibinfo  {journal} {Phys.
  Rev. B}\ }\textbf {\bibinfo {volume} {89}},\ \bibinfo {pages} {125309}
  (\bibinfo {year} {2014})}\BibitemShut {NoStop}%
\bibitem [{\citenamefont {Richter}(2017)}]{richter2017nanoplatelets}%
  \BibitemOpen
  \bibfield  {author} {\bibinfo {author} {\bibfnamefont {M.}~\bibnamefont
  {Richter}},\ }\href@noop {} {\bibfield  {journal} {\bibinfo  {journal}
  {Physical Review Materials}\ }\textbf {\bibinfo {volume} {1}},\ \bibinfo
  {pages} {016001} (\bibinfo {year} {2017})}\BibitemShut {NoStop}%
\bibitem [{\citenamefont {Singh}\ \emph {et~al.}(2017)\citenamefont {Singh},
  \citenamefont {Richter}, \citenamefont {Moody}, \citenamefont {Siemens},
  \citenamefont {Li},\ and\ \citenamefont {Cundiff}}]{PhysRevB.95.235307}%
  \BibitemOpen
  \bibfield  {author} {\bibinfo {author} {\bibfnamefont {R.}~\bibnamefont
  {Singh}}, \bibinfo {author} {\bibfnamefont {M.}~\bibnamefont {Richter}},
  \bibinfo {author} {\bibfnamefont {G.}~\bibnamefont {Moody}}, \bibinfo
  {author} {\bibfnamefont {M.~E.}\ \bibnamefont {Siemens}}, \bibinfo {author}
  {\bibfnamefont {H.}~\bibnamefont {Li}}, \ and\ \bibinfo {author}
  {\bibfnamefont {S.~T.}\ \bibnamefont {Cundiff}},\ }\href {\doibase
  10.1103/PhysRevB.95.235307} {\bibfield  {journal} {\bibinfo  {journal} {Phys.
  Rev. B}\ }\textbf {\bibinfo {volume} {95}},\ \bibinfo {pages} {235307}
  (\bibinfo {year} {2017})}\BibitemShut {NoStop}%
\bibitem [{\citenamefont {H{\"u}ser}\ \emph {et~al.}(2013)\citenamefont
  {H{\"u}ser}, \citenamefont {Olsen},\ and\ \citenamefont
  {Thygesen}}]{huser2013dielectric}%
  \BibitemOpen
  \bibfield  {author} {\bibinfo {author} {\bibfnamefont {F.}~\bibnamefont
  {H{\"u}ser}}, \bibinfo {author} {\bibfnamefont {T.}~\bibnamefont {Olsen}}, \
  and\ \bibinfo {author} {\bibfnamefont {K.~S.}\ \bibnamefont {Thygesen}},\
  }\href@noop {} {\bibfield  {journal} {\bibinfo  {journal} {Physical Review
  B}\ }\textbf {\bibinfo {volume} {88}},\ \bibinfo {pages} {245309} (\bibinfo
  {year} {2013})}\BibitemShut {NoStop}%
\bibitem [{\citenamefont {Rinke}\ \emph {et~al.}(2012)\citenamefont {Rinke},
  \citenamefont {Schleife}, \citenamefont {Kioupakis}, \citenamefont {Janotti},
  \citenamefont {R\"odl}, \citenamefont {Bechstedt}, \citenamefont
  {Scheffler},\ and\ \citenamefont {Van~de Walle}}]{PhysRevLett.108.126404}%
  \BibitemOpen
  \bibfield  {author} {\bibinfo {author} {\bibfnamefont {P.}~\bibnamefont
  {Rinke}}, \bibinfo {author} {\bibfnamefont {A.}~\bibnamefont {Schleife}},
  \bibinfo {author} {\bibfnamefont {E.}~\bibnamefont {Kioupakis}}, \bibinfo
  {author} {\bibfnamefont {A.}~\bibnamefont {Janotti}}, \bibinfo {author}
  {\bibfnamefont {C.}~\bibnamefont {R\"odl}}, \bibinfo {author} {\bibfnamefont
  {F.}~\bibnamefont {Bechstedt}}, \bibinfo {author} {\bibfnamefont
  {M.}~\bibnamefont {Scheffler}}, \ and\ \bibinfo {author} {\bibfnamefont
  {C.~G.}\ \bibnamefont {Van~de Walle}},\ }\href {\doibase
  10.1103/PhysRevLett.108.126404} {\bibfield  {journal} {\bibinfo  {journal}
  {Phys. Rev. Lett.}\ }\textbf {\bibinfo {volume} {108}},\ \bibinfo {pages}
  {126404} (\bibinfo {year} {2012})}\BibitemShut {NoStop}%
\bibitem [{\citenamefont {Cudazzo}\ \emph {et~al.}(2010)\citenamefont
  {Cudazzo}, \citenamefont {Attaccalite}, \citenamefont {Tokatly},\ and\
  \citenamefont {Rubio}}]{PhysRevLett.104.226804}%
  \BibitemOpen
  \bibfield  {author} {\bibinfo {author} {\bibfnamefont {P.}~\bibnamefont
  {Cudazzo}}, \bibinfo {author} {\bibfnamefont {C.}~\bibnamefont
  {Attaccalite}}, \bibinfo {author} {\bibfnamefont {I.~V.}\ \bibnamefont
  {Tokatly}}, \ and\ \bibinfo {author} {\bibfnamefont {A.}~\bibnamefont
  {Rubio}},\ }\href {\doibase 10.1103/PhysRevLett.104.226804} {\bibfield
  {journal} {\bibinfo  {journal} {Phys. Rev. Lett.}\ }\textbf {\bibinfo
  {volume} {104}},\ \bibinfo {pages} {226804} (\bibinfo {year}
  {2010})}\BibitemShut {NoStop}%
\bibitem [{\citenamefont {Qiu}\ \emph {et~al.}(2013)\citenamefont {Qiu},
  \citenamefont {Felipe},\ and\ \citenamefont {Louie}}]{qiu2013optical}%
  \BibitemOpen
  \bibfield  {author} {\bibinfo {author} {\bibfnamefont {D.~Y.}\ \bibnamefont
  {Qiu}}, \bibinfo {author} {\bibfnamefont {H.}~\bibnamefont {Felipe}}, \ and\
  \bibinfo {author} {\bibfnamefont {S.~G.}\ \bibnamefont {Louie}},\ }\href@noop
  {} {\bibfield  {journal} {\bibinfo  {journal} {Phys. Rev. Lett.}\ }\textbf
  {\bibinfo {volume} {111}},\ \bibinfo {pages} {216805} (\bibinfo {year}
  {2013})}\BibitemShut {NoStop}%
\bibitem [{\citenamefont {Laskowski}\ \emph {et~al.}(2005)\citenamefont
  {Laskowski}, \citenamefont {Christensen}, \citenamefont {Santi},\ and\
  \citenamefont {Ambrosch-Draxl}}]{laskowski2005ab}%
  \BibitemOpen
  \bibfield  {author} {\bibinfo {author} {\bibfnamefont {R.}~\bibnamefont
  {Laskowski}}, \bibinfo {author} {\bibfnamefont {N.~E.}\ \bibnamefont
  {Christensen}}, \bibinfo {author} {\bibfnamefont {G.}~\bibnamefont {Santi}},
  \ and\ \bibinfo {author} {\bibfnamefont {C.}~\bibnamefont {Ambrosch-Draxl}},\
  }\href@noop {} {\bibfield  {journal} {\bibinfo  {journal} {Physical Review
  B}\ }\textbf {\bibinfo {volume} {72}},\ \bibinfo {pages} {035204} (\bibinfo
  {year} {2005})}\BibitemShut {NoStop}%
\bibitem [{\citenamefont {Qiu}\ \emph {et~al.}(2015)\citenamefont {Qiu},
  \citenamefont {Cao},\ and\ \citenamefont {Louie}}]{qiu2015nonanalyticity}%
  \BibitemOpen
  \bibfield  {author} {\bibinfo {author} {\bibfnamefont {D.~Y.}\ \bibnamefont
  {Qiu}}, \bibinfo {author} {\bibfnamefont {T.}~\bibnamefont {Cao}}, \ and\
  \bibinfo {author} {\bibfnamefont {S.~G.}\ \bibnamefont {Louie}},\ }\href@noop
  {} {\bibfield  {journal} {\bibinfo  {journal} {Physical review letters}\
  }\textbf {\bibinfo {volume} {115}},\ \bibinfo {pages} {176801} (\bibinfo
  {year} {2015})}\BibitemShut {NoStop}%
\bibitem [{\citenamefont {Torun}\ \emph {et~al.}(2018)\citenamefont {Torun},
  \citenamefont {Miranda}, \citenamefont {Molina-S\'anchez},\ and\
  \citenamefont {Wirtz}}]{PhysRevB.97.245427}%
  \BibitemOpen
  \bibfield  {author} {\bibinfo {author} {\bibfnamefont {E.}~\bibnamefont
  {Torun}}, \bibinfo {author} {\bibfnamefont {H.~P.~C.}\ \bibnamefont
  {Miranda}}, \bibinfo {author} {\bibfnamefont {A.}~\bibnamefont
  {Molina-S\'anchez}}, \ and\ \bibinfo {author} {\bibfnamefont
  {L.}~\bibnamefont {Wirtz}},\ }\href {\doibase 10.1103/PhysRevB.97.245427}
  {\bibfield  {journal} {\bibinfo  {journal} {Phys. Rev. B}\ }\textbf {\bibinfo
  {volume} {97}},\ \bibinfo {pages} {245427} (\bibinfo {year}
  {2018})}\BibitemShut {NoStop}%
\bibitem [{\citenamefont {St{\'e}b{\'e}}\ \emph {et~al.}(1997)\citenamefont
  {St{\'e}b{\'e}}, \citenamefont {Munschy}, \citenamefont {Stauffer},
  \citenamefont {Dujardin},\ and\ \citenamefont {Murat}}]{stebe1997excitonic}%
  \BibitemOpen
  \bibfield  {author} {\bibinfo {author} {\bibfnamefont {B.}~\bibnamefont
  {St{\'e}b{\'e}}}, \bibinfo {author} {\bibfnamefont {G.}~\bibnamefont
  {Munschy}}, \bibinfo {author} {\bibfnamefont {L.}~\bibnamefont {Stauffer}},
  \bibinfo {author} {\bibfnamefont {F.}~\bibnamefont {Dujardin}}, \ and\
  \bibinfo {author} {\bibfnamefont {J.}~\bibnamefont {Murat}},\ }\href@noop {}
  {\bibfield  {journal} {\bibinfo  {journal} {Physical Review B}\ }\textbf
  {\bibinfo {volume} {56}},\ \bibinfo {pages} {12454} (\bibinfo {year}
  {1997})}\BibitemShut {NoStop}%
\bibitem [{\citenamefont {Mayrock}\ \emph {et~al.}(1999)\citenamefont
  {Mayrock}, \citenamefont {W{\"u}nsche}, \citenamefont {Henneberger},
  \citenamefont {Riva}, \citenamefont {Schweigert},\ and\ \citenamefont
  {Peeters}}]{mayrock1999weak}%
  \BibitemOpen
  \bibfield  {author} {\bibinfo {author} {\bibfnamefont {O.}~\bibnamefont
  {Mayrock}}, \bibinfo {author} {\bibfnamefont {H.-J.}\ \bibnamefont
  {W{\"u}nsche}}, \bibinfo {author} {\bibfnamefont {F.}~\bibnamefont
  {Henneberger}}, \bibinfo {author} {\bibfnamefont {C.}~\bibnamefont {Riva}},
  \bibinfo {author} {\bibfnamefont {V.}~\bibnamefont {Schweigert}}, \ and\
  \bibinfo {author} {\bibfnamefont {F.}~\bibnamefont {Peeters}},\ }\href@noop
  {} {\bibfield  {journal} {\bibinfo  {journal} {Physical Review B}\ }\textbf
  {\bibinfo {volume} {60}},\ \bibinfo {pages} {5582} (\bibinfo {year}
  {1999})}\BibitemShut {NoStop}%
\bibitem [{\citenamefont {Singh}\ \emph {et~al.}(2016)\citenamefont {Singh},
  \citenamefont {Moody}, \citenamefont {Tran}, \citenamefont {Scott},
  \citenamefont {Overbeck}, \citenamefont {Bergh\"auser}, \citenamefont
  {Schaibley}, \citenamefont {Seifert}, \citenamefont {Pleskot}, \citenamefont
  {Gabor}, \citenamefont {Yan}, \citenamefont {Mandrus}, \citenamefont
  {Richter}, \citenamefont {Malic}, \citenamefont {Xu},\ and\ \citenamefont
  {Li}}]{PhysRevB.93.041401}%
  \BibitemOpen
  \bibfield  {author} {\bibinfo {author} {\bibfnamefont {A.}~\bibnamefont
  {Singh}}, \bibinfo {author} {\bibfnamefont {G.}~\bibnamefont {Moody}},
  \bibinfo {author} {\bibfnamefont {K.}~\bibnamefont {Tran}}, \bibinfo {author}
  {\bibfnamefont {M.~E.}\ \bibnamefont {Scott}}, \bibinfo {author}
  {\bibfnamefont {V.}~\bibnamefont {Overbeck}}, \bibinfo {author}
  {\bibfnamefont {G.}~\bibnamefont {Bergh\"auser}}, \bibinfo {author}
  {\bibfnamefont {J.}~\bibnamefont {Schaibley}}, \bibinfo {author}
  {\bibfnamefont {E.~J.}\ \bibnamefont {Seifert}}, \bibinfo {author}
  {\bibfnamefont {D.}~\bibnamefont {Pleskot}}, \bibinfo {author} {\bibfnamefont
  {N.~M.}\ \bibnamefont {Gabor}}, \bibinfo {author} {\bibfnamefont
  {J.}~\bibnamefont {Yan}}, \bibinfo {author} {\bibfnamefont {D.~G.}\
  \bibnamefont {Mandrus}}, \bibinfo {author} {\bibfnamefont {M.}~\bibnamefont
  {Richter}}, \bibinfo {author} {\bibfnamefont {E.}~\bibnamefont {Malic}},
  \bibinfo {author} {\bibfnamefont {X.}~\bibnamefont {Xu}}, \ and\ \bibinfo
  {author} {\bibfnamefont {X.}~\bibnamefont {Li}},\ }\href {\doibase
  10.1103/PhysRevB.93.041401} {\bibfield  {journal} {\bibinfo  {journal} {Phys.
  Rev. B}\ }\textbf {\bibinfo {volume} {93}},\ \bibinfo {pages} {041401}
  (\bibinfo {year} {2016})}\BibitemShut {NoStop}%
\bibitem [{\citenamefont {Florian}\ \emph {et~al.}(2018)\citenamefont
  {Florian}, \citenamefont {Hartmann}, \citenamefont {Steinhoff}, \citenamefont
  {Klein}, \citenamefont {Holleitner}, \citenamefont {Finley}, \citenamefont
  {Wehling}, \citenamefont {Kaniber},\ and\ \citenamefont
  {Gies}}]{10.1021acs.nanolett.8b00840}%
  \BibitemOpen
  \bibfield  {author} {\bibinfo {author} {\bibfnamefont {M.}~\bibnamefont
  {Florian}}, \bibinfo {author} {\bibfnamefont {M.}~\bibnamefont {Hartmann}},
  \bibinfo {author} {\bibfnamefont {A.}~\bibnamefont {Steinhoff}}, \bibinfo
  {author} {\bibfnamefont {J.}~\bibnamefont {Klein}}, \bibinfo {author}
  {\bibfnamefont {A.~W.}\ \bibnamefont {Holleitner}}, \bibinfo {author}
  {\bibfnamefont {J.~J.}\ \bibnamefont {Finley}}, \bibinfo {author}
  {\bibfnamefont {T.~O.}\ \bibnamefont {Wehling}}, \bibinfo {author}
  {\bibfnamefont {M.}~\bibnamefont {Kaniber}}, \ and\ \bibinfo {author}
  {\bibfnamefont {C.}~\bibnamefont {Gies}},\ }\href {\doibase
  10.1021/acs.nanolett.8b00840} {\bibfield  {journal} {\bibinfo  {journal}
  {Nano Letters}\ }\textbf {\bibinfo {volume} {18}},\ \bibinfo {pages} {2725}
  (\bibinfo {year} {2018})},\ \bibinfo {note} {pMID: 29558797},\ \Eprint
  {http://arxiv.org/abs/https://doi.org/10.1021/acs.nanolett.8b00840}
  {https://doi.org/10.1021/acs.nanolett.8b00840} \BibitemShut {NoStop}%
\bibitem [{\citenamefont {Drüppel}\ \emph {et~al.}(2017)\citenamefont
  {Drüppel}, \citenamefont {Deilmann}, \citenamefont {Krüger},\ and\
  \citenamefont {Rohlfing}}]{Drueppel2017}%
  \BibitemOpen
  \bibfield  {author} {\bibinfo {author} {\bibfnamefont {M.}~\bibnamefont
  {Drüppel}}, \bibinfo {author} {\bibfnamefont {T.}~\bibnamefont {Deilmann}},
  \bibinfo {author} {\bibfnamefont {P.}~\bibnamefont {Krüger}}, \ and\
  \bibinfo {author} {\bibfnamefont {M.}~\bibnamefont {Rohlfing}},\ }\href@noop
  {} {\bibfield  {journal} {\bibinfo  {journal} {Nature Communications}\
  }\textbf {\bibinfo {volume} {8}},\ \bibinfo {pages} {2117} (\bibinfo {year}
  {2017})}\BibitemShut {NoStop}%
\bibitem [{\citenamefont {Kabuss}\ \emph {et~al.}(2011)\citenamefont {Kabuss},
  \citenamefont {Carmele}, \citenamefont {Richter},\ and\ \citenamefont
  {Knorr}}]{PhysRevB.84.125324}%
  \BibitemOpen
  \bibfield  {author} {\bibinfo {author} {\bibfnamefont {J.}~\bibnamefont
  {Kabuss}}, \bibinfo {author} {\bibfnamefont {A.}~\bibnamefont {Carmele}},
  \bibinfo {author} {\bibfnamefont {M.}~\bibnamefont {Richter}}, \ and\
  \bibinfo {author} {\bibfnamefont {A.}~\bibnamefont {Knorr}},\ }\href
  {\doibase 10.1103/PhysRevB.84.125324} {\bibfield  {journal} {\bibinfo
  {journal} {Phys. Rev. B}\ }\textbf {\bibinfo {volume} {84}},\ \bibinfo
  {pages} {125324} (\bibinfo {year} {2011})}\BibitemShut {NoStop}%
\bibitem [{\citenamefont {Carmele}\ \emph {et~al.}(2010)\citenamefont
  {Carmele}, \citenamefont {Richter}, \citenamefont {Chow},\ and\ \citenamefont
  {Knorr}}]{PhysRevLett.104.156801}%
  \BibitemOpen
  \bibfield  {author} {\bibinfo {author} {\bibfnamefont {A.}~\bibnamefont
  {Carmele}}, \bibinfo {author} {\bibfnamefont {M.}~\bibnamefont {Richter}},
  \bibinfo {author} {\bibfnamefont {W.~W.}\ \bibnamefont {Chow}}, \ and\
  \bibinfo {author} {\bibfnamefont {A.}~\bibnamefont {Knorr}},\ }\href
  {\doibase 10.1103/PhysRevLett.104.156801} {\bibfield  {journal} {\bibinfo
  {journal} {Phys. Rev. Lett.}\ }\textbf {\bibinfo {volume} {104}},\ \bibinfo
  {pages} {156801} (\bibinfo {year} {2010})}\BibitemShut {NoStop}%
\bibitem [{\citenamefont {Leymann}\ \emph {et~al.}(2014)\citenamefont
  {Leymann}, \citenamefont {Foerster},\ and\ \citenamefont
  {Wiersig}}]{PhysRevB.89.085308}%
  \BibitemOpen
  \bibfield  {author} {\bibinfo {author} {\bibfnamefont {H.~A.~M.}\
  \bibnamefont {Leymann}}, \bibinfo {author} {\bibfnamefont {A.}~\bibnamefont
  {Foerster}}, \ and\ \bibinfo {author} {\bibfnamefont {J.}~\bibnamefont
  {Wiersig}},\ }\href {\doibase 10.1103/PhysRevB.89.085308} {\bibfield
  {journal} {\bibinfo  {journal} {Phys. Rev. B}\ }\textbf {\bibinfo {volume}
  {89}},\ \bibinfo {pages} {085308} (\bibinfo {year} {2014})}\BibitemShut
  {NoStop}%
\bibitem [{\citenamefont {Gegg}\ and\ \citenamefont
  {Richter}(2016)}]{gegg2016}%
  \BibitemOpen
  \bibfield  {author} {\bibinfo {author} {\bibfnamefont {M.}~\bibnamefont
  {Gegg}}\ and\ \bibinfo {author} {\bibfnamefont {M.}~\bibnamefont {Richter}},\
  }\href {http://stacks.iop.org/1367-2630/18/i=4/a=043037} {\bibfield
  {journal} {\bibinfo  {journal} {New Journal of Physics}\ }\textbf {\bibinfo
  {volume} {18}},\ \bibinfo {pages} {043037} (\bibinfo {year}
  {2016})}\BibitemShut {NoStop}%
\bibitem [{\citenamefont {Gegg}\ and\ \citenamefont
  {Richter}(2017)}]{Gegg2017}%
  \BibitemOpen
  \bibfield  {author} {\bibinfo {author} {\bibfnamefont {M.}~\bibnamefont
  {Gegg}}\ and\ \bibinfo {author} {\bibfnamefont {M.}~\bibnamefont {Richter}},\
  }\href {\doibase https://doi.org/10.1038/s41598-017-16178-8} {\bibfield
  {journal} {\bibinfo  {journal} {Scientific Reports}\ }\textbf {\bibinfo
  {volume} {7}},\ \bibinfo {pages} {16304} (\bibinfo {year}
  {2017})}\BibitemShut {NoStop}%
\bibitem [{\citenamefont {Verstraete}\ and\ \citenamefont
  {Cirac}(2006)}]{verstraete2006matrix}%
  \BibitemOpen
  \bibfield  {author} {\bibinfo {author} {\bibfnamefont {F.}~\bibnamefont
  {Verstraete}}\ and\ \bibinfo {author} {\bibfnamefont {J.~I.}\ \bibnamefont
  {Cirac}},\ }\href@noop {} {\bibfield  {journal} {\bibinfo  {journal}
  {Physical Review B}\ }\textbf {\bibinfo {volume} {73}},\ \bibinfo {pages}
  {094423} (\bibinfo {year} {2006})}\BibitemShut {NoStop}%
\bibitem [{\citenamefont {Vidal}(2007)}]{vidal2007classical}%
  \BibitemOpen
  \bibfield  {author} {\bibinfo {author} {\bibfnamefont {G.}~\bibnamefont
  {Vidal}},\ }\href@noop {} {\bibfield  {journal} {\bibinfo  {journal}
  {Physical review letters}\ }\textbf {\bibinfo {volume} {98}},\ \bibinfo
  {pages} {070201} (\bibinfo {year} {2007})}\BibitemShut {NoStop}%
\bibitem [{\citenamefont {Cirac}\ \emph {et~al.}(2017)\citenamefont {Cirac},
  \citenamefont {Pérez-García}, \citenamefont {Schuch},\ and\ \citenamefont
  {Verstraete}}]{CIRAC2017100}%
  \BibitemOpen
  \bibfield  {author} {\bibinfo {author} {\bibfnamefont {J.}~\bibnamefont
  {Cirac}}, \bibinfo {author} {\bibfnamefont {D.}~\bibnamefont
  {Pérez-García}}, \bibinfo {author} {\bibfnamefont {N.}~\bibnamefont
  {Schuch}}, \ and\ \bibinfo {author} {\bibfnamefont {F.}~\bibnamefont
  {Verstraete}},\ }\href {\doibase https://doi.org/10.1016/j.aop.2016.12.030}
  {\bibfield  {journal} {\bibinfo  {journal} {Annals of Physics}\ }\textbf
  {\bibinfo {volume} {378}},\ \bibinfo {pages} {100 } (\bibinfo {year}
  {2017})}\BibitemShut {NoStop}%
\bibitem [{\citenamefont {Clark}\ \emph {et~al.}(2010)\citenamefont {Clark},
  \citenamefont {Prior}, \citenamefont {Hartmann}, \citenamefont {Jaksch},\
  and\ \citenamefont {Plenio}}]{plenioheisenberg}%
  \BibitemOpen
  \bibfield  {author} {\bibinfo {author} {\bibfnamefont {S.~R.}\ \bibnamefont
  {Clark}}, \bibinfo {author} {\bibfnamefont {J.}~\bibnamefont {Prior}},
  \bibinfo {author} {\bibfnamefont {M.~J.}\ \bibnamefont {Hartmann}}, \bibinfo
  {author} {\bibfnamefont {D.}~\bibnamefont {Jaksch}}, \ and\ \bibinfo {author}
  {\bibfnamefont {M.~B.}\ \bibnamefont {Plenio}},\ }\href
  {http://stacks.iop.org/1367-2630/12/i=2/a=025005} {\bibfield  {journal}
  {\bibinfo  {journal} {New Journal of Physics}\ }\textbf {\bibinfo {volume}
  {12}},\ \bibinfo {pages} {025005} (\bibinfo {year} {2010})}\BibitemShut
  {NoStop}%
\bibitem [{\citenamefont {Werner}\ \emph {et~al.}(2016)\citenamefont {Werner},
  \citenamefont {Jaschke}, \citenamefont {Silvi}, \citenamefont {Kliesch},
  \citenamefont {Calarco}, \citenamefont {Eisert},\ and\ \citenamefont
  {Montangero}}]{PhysRevLett.116.237201}%
  \BibitemOpen
  \bibfield  {author} {\bibinfo {author} {\bibfnamefont {A.~H.}\ \bibnamefont
  {Werner}}, \bibinfo {author} {\bibfnamefont {D.}~\bibnamefont {Jaschke}},
  \bibinfo {author} {\bibfnamefont {P.}~\bibnamefont {Silvi}}, \bibinfo
  {author} {\bibfnamefont {M.}~\bibnamefont {Kliesch}}, \bibinfo {author}
  {\bibfnamefont {T.}~\bibnamefont {Calarco}}, \bibinfo {author} {\bibfnamefont
  {J.}~\bibnamefont {Eisert}}, \ and\ \bibinfo {author} {\bibfnamefont
  {S.}~\bibnamefont {Montangero}},\ }\href {\doibase
  10.1103/PhysRevLett.116.237201} {\bibfield  {journal} {\bibinfo  {journal}
  {Phys. Rev. Lett.}\ }\textbf {\bibinfo {volume} {116}},\ \bibinfo {pages}
  {237201} (\bibinfo {year} {2016})}\BibitemShut {NoStop}%
\bibitem [{\citenamefont {Shi}\ \emph {et~al.}(2018)\citenamefont {Shi},
  \citenamefont {Xu}, \citenamefont {Yan},\ and\ \citenamefont {Xu}}]{shi2018}%
  \BibitemOpen
  \bibfield  {author} {\bibinfo {author} {\bibfnamefont {Q.}~\bibnamefont
  {Shi}}, \bibinfo {author} {\bibfnamefont {Y.}~\bibnamefont {Xu}}, \bibinfo
  {author} {\bibfnamefont {Y.}~\bibnamefont {Yan}}, \ and\ \bibinfo {author}
  {\bibfnamefont {M.}~\bibnamefont {Xu}},\ }\href {\doibase 10.1063/1.5026753}
  {\bibfield  {journal} {\bibinfo  {journal} {The Journal of Chemical Physics}\
  }\textbf {\bibinfo {volume} {148}},\ \bibinfo {pages} {174102} (\bibinfo
  {year} {2018})},\ \Eprint
  {http://arxiv.org/abs/https://doi.org/10.1063/1.5026753}
  {https://doi.org/10.1063/1.5026753} \BibitemShut {NoStop}%
\bibitem [{\citenamefont {Oseledets}(2009)}]{oseledets2009approximation}%
  \BibitemOpen
  \bibfield  {author} {\bibinfo {author} {\bibfnamefont {I.}~\bibnamefont
  {Oseledets}},\ }in\ \href@noop {} {\emph {\bibinfo {booktitle} {Doklady
  Mathematics}}},\ Vol.~\bibinfo {volume} {80}\ (\bibinfo {organization}
  {Springer},\ \bibinfo {year} {2009})\ pp.\ \bibinfo {pages}
  {653--654}\BibitemShut {NoStop}%
\bibitem [{\citenamefont {Oseledets}(2010)}]{oseledets2010approximation}%
  \BibitemOpen
  \bibfield  {author} {\bibinfo {author} {\bibfnamefont {I.~V.}\ \bibnamefont
  {Oseledets}},\ }\href@noop {} {\bibfield  {journal} {\bibinfo  {journal}
  {SIAM Journal on Matrix Analysis and Applications}\ }\textbf {\bibinfo
  {volume} {31}},\ \bibinfo {pages} {2130} (\bibinfo {year}
  {2010})}\BibitemShut {NoStop}%
\bibitem [{\citenamefont {Khoromskij}(2011)}]{khoromskij2011dlog}%
  \BibitemOpen
  \bibfield  {author} {\bibinfo {author} {\bibfnamefont {B.~N.}\ \bibnamefont
  {Khoromskij}},\ }\href@noop {} {\bibfield  {journal} {\bibinfo  {journal}
  {Constructive Approximation}\ }\textbf {\bibinfo {volume} {34}},\ \bibinfo
  {pages} {257} (\bibinfo {year} {2011})}\BibitemShut {NoStop}%
\bibitem [{\citenamefont {Kazeev}\ and\ \citenamefont
  {Khoromskij}(2012)}]{kazeev2012low}%
  \BibitemOpen
  \bibfield  {author} {\bibinfo {author} {\bibfnamefont {V.~A.}\ \bibnamefont
  {Kazeev}}\ and\ \bibinfo {author} {\bibfnamefont {B.~N.}\ \bibnamefont
  {Khoromskij}},\ }\href@noop {} {\bibfield  {journal} {\bibinfo  {journal}
  {SIAM J. Matrix Anal. Appl.}\ }\textbf {\bibinfo {volume} {33}},\ \bibinfo
  {pages} {742} (\bibinfo {year} {2012})}\BibitemShut {NoStop}%
\bibitem [{\citenamefont {Khoromskaia}\ and\ \citenamefont
  {Khoromskij}(2015)}]{khoromskaia2015tensor}%
  \BibitemOpen
  \bibfield  {author} {\bibinfo {author} {\bibfnamefont {V.}~\bibnamefont
  {Khoromskaia}}\ and\ \bibinfo {author} {\bibfnamefont {B.~N.}\ \bibnamefont
  {Khoromskij}},\ }\href@noop {} {\bibfield  {journal} {\bibinfo  {journal}
  {Physical Chemistry Chemical Physics}\ }\textbf {\bibinfo {volume} {17}},\
  \bibinfo {pages} {31491} (\bibinfo {year} {2015})}\BibitemShut {NoStop}%
\bibitem [{\citenamefont {Benner}\ \emph {et~al.}(2017)\citenamefont {Benner},
  \citenamefont {Dolgov}, \citenamefont {Khoromskaia},\ and\ \citenamefont
  {Khoromskij}}]{benner2017fast}%
  \BibitemOpen
  \bibfield  {author} {\bibinfo {author} {\bibfnamefont {P.}~\bibnamefont
  {Benner}}, \bibinfo {author} {\bibfnamefont {S.}~\bibnamefont {Dolgov}},
  \bibinfo {author} {\bibfnamefont {V.}~\bibnamefont {Khoromskaia}}, \ and\
  \bibinfo {author} {\bibfnamefont {B.~N.}\ \bibnamefont {Khoromskij}},\
  }\href@noop {} {\bibfield  {journal} {\bibinfo  {journal} {Journal of
  Computational Physics}\ }\textbf {\bibinfo {volume} {334}},\ \bibinfo {pages}
  {221} (\bibinfo {year} {2017})}\BibitemShut {NoStop}%
\bibitem [{\citenamefont {Hao}\ \emph {et~al.}(2017)\citenamefont {Hao},
  \citenamefont {Specht}, \citenamefont {Nagler}, \citenamefont {Xu},
  \citenamefont {Tran}, \citenamefont {Singh}, \citenamefont {Dass},
  \citenamefont {Sch{\"u}ller}, \citenamefont {Korn}, \citenamefont {Richter}
  \emph {et~al.}}]{hao2017neutral}%
  \BibitemOpen
  \bibfield  {author} {\bibinfo {author} {\bibfnamefont {K.}~\bibnamefont
  {Hao}}, \bibinfo {author} {\bibfnamefont {J.~F.}\ \bibnamefont {Specht}},
  \bibinfo {author} {\bibfnamefont {P.}~\bibnamefont {Nagler}}, \bibinfo
  {author} {\bibfnamefont {L.}~\bibnamefont {Xu}}, \bibinfo {author}
  {\bibfnamefont {K.}~\bibnamefont {Tran}}, \bibinfo {author} {\bibfnamefont
  {A.}~\bibnamefont {Singh}}, \bibinfo {author} {\bibfnamefont {C.~K.}\
  \bibnamefont {Dass}}, \bibinfo {author} {\bibfnamefont {C.}~\bibnamefont
  {Sch{\"u}ller}}, \bibinfo {author} {\bibfnamefont {T.}~\bibnamefont {Korn}},
  \bibinfo {author} {\bibfnamefont {M.}~\bibnamefont {Richter}},  \emph
  {et~al.},\ }\href@noop {} {\bibfield  {journal} {\bibinfo  {journal} {Nature
  communications}\ }\textbf {\bibinfo {volume} {8}},\ \bibinfo {pages} {15552}
  (\bibinfo {year} {2017})}\BibitemShut {NoStop}%
\bibitem [{\citenamefont {Zhang}\ \emph {et~al.}(2015)\citenamefont {Zhang},
  \citenamefont {You}, \citenamefont {Zhao},\ and\ \citenamefont
  {Heinz}}]{zhang2015experimental}%
  \BibitemOpen
  \bibfield  {author} {\bibinfo {author} {\bibfnamefont {X.-X.}\ \bibnamefont
  {Zhang}}, \bibinfo {author} {\bibfnamefont {Y.}~\bibnamefont {You}}, \bibinfo
  {author} {\bibfnamefont {S.~Y.~F.}\ \bibnamefont {Zhao}}, \ and\ \bibinfo
  {author} {\bibfnamefont {T.~F.}\ \bibnamefont {Heinz}},\ }\href@noop {}
  {\bibfield  {journal} {\bibinfo  {journal} {Phys. Rev. Lett.}\ }\textbf
  {\bibinfo {volume} {115}},\ \bibinfo {pages} {257403} (\bibinfo {year}
  {2015})}\BibitemShut {NoStop}%
\bibitem [{\citenamefont {Sie}\ \emph {et~al.}(2015)\citenamefont {Sie},
  \citenamefont {Frenzel}, \citenamefont {Lee}, \citenamefont {Kong},\ and\
  \citenamefont {Gedik}}]{sie2015intervalley}%
  \BibitemOpen
  \bibfield  {author} {\bibinfo {author} {\bibfnamefont {E.~J.}\ \bibnamefont
  {Sie}}, \bibinfo {author} {\bibfnamefont {A.~J.}\ \bibnamefont {Frenzel}},
  \bibinfo {author} {\bibfnamefont {Y.-H.}\ \bibnamefont {Lee}}, \bibinfo
  {author} {\bibfnamefont {J.}~\bibnamefont {Kong}}, \ and\ \bibinfo {author}
  {\bibfnamefont {N.}~\bibnamefont {Gedik}},\ }\href@noop {} {\bibfield
  {journal} {\bibinfo  {journal} {Phys. Rev. B}\ }\textbf {\bibinfo {volume}
  {92}},\ \bibinfo {pages} {125417} (\bibinfo {year} {2015})}\BibitemShut
  {NoStop}%
\bibitem [{\citenamefont {Olsen}\ \emph {et~al.}(2016)\citenamefont {Olsen},
  \citenamefont {Latini}, \citenamefont {Rasmussen},\ and\ \citenamefont
  {Thygesen}}]{olsen2016simple}%
  \BibitemOpen
  \bibfield  {author} {\bibinfo {author} {\bibfnamefont {T.}~\bibnamefont
  {Olsen}}, \bibinfo {author} {\bibfnamefont {S.}~\bibnamefont {Latini}},
  \bibinfo {author} {\bibfnamefont {F.}~\bibnamefont {Rasmussen}}, \ and\
  \bibinfo {author} {\bibfnamefont {K.~S.}\ \bibnamefont {Thygesen}},\
  }\href@noop {} {\bibfield  {journal} {\bibinfo  {journal} {Physical review
  letters}\ }\textbf {\bibinfo {volume} {116}},\ \bibinfo {pages} {056401}
  (\bibinfo {year} {2016})}\BibitemShut {NoStop}%
\bibitem [{\citenamefont {Szyniszewski}\ \emph {et~al.}(2017)\citenamefont
  {Szyniszewski}, \citenamefont {Mostaani}, \citenamefont {Drummond},\ and\
  \citenamefont {Fal'ko}}]{PhysRevB.95.081301}%
  \BibitemOpen
  \bibfield  {author} {\bibinfo {author} {\bibfnamefont {M.}~\bibnamefont
  {Szyniszewski}}, \bibinfo {author} {\bibfnamefont {E.}~\bibnamefont
  {Mostaani}}, \bibinfo {author} {\bibfnamefont {N.~D.}\ \bibnamefont
  {Drummond}}, \ and\ \bibinfo {author} {\bibfnamefont {V.~I.}\ \bibnamefont
  {Fal'ko}},\ }\href {\doibase 10.1103/PhysRevB.95.081301} {\bibfield
  {journal} {\bibinfo  {journal} {Phys. Rev. B}\ }\textbf {\bibinfo {volume}
  {95}},\ \bibinfo {pages} {081301} (\bibinfo {year} {2017})}\BibitemShut
  {NoStop}%
\bibitem [{\citenamefont {Selig}\ \emph {et~al.}(2018)\citenamefont {Selig},
  \citenamefont {Bergh{\"a}user}, \citenamefont {Richter}, \citenamefont
  {Bratschitsch}, \citenamefont {Knorr},\ and\ \citenamefont
  {Malic}}]{maltedarkbright}%
  \BibitemOpen
  \bibfield  {author} {\bibinfo {author} {\bibfnamefont {M.}~\bibnamefont
  {Selig}}, \bibinfo {author} {\bibfnamefont {G.}~\bibnamefont
  {Bergh{\"a}user}}, \bibinfo {author} {\bibfnamefont {M.}~\bibnamefont
  {Richter}}, \bibinfo {author} {\bibfnamefont {R.}~\bibnamefont
  {Bratschitsch}}, \bibinfo {author} {\bibfnamefont {A.}~\bibnamefont {Knorr}},
  \ and\ \bibinfo {author} {\bibfnamefont {E.}~\bibnamefont {Malic}},\ }\href
  {http://stacks.iop.org/2053-1583/5/i=3/a=035017} {\bibfield  {journal}
  {\bibinfo  {journal} {2D Materials}\ }\textbf {\bibinfo {volume} {5}},\
  \bibinfo {pages} {035017} (\bibinfo {year} {2018})}\BibitemShut {NoStop}%
\bibitem [{\citenamefont {Ridolfi}\ \emph {et~al.}(2015)\citenamefont
  {Ridolfi}, \citenamefont {Le}, \citenamefont {Rahman}, \citenamefont
  {Mucciolo},\ and\ \citenamefont {Lewenkopf}}]{ridolfi2015tight}%
  \BibitemOpen
  \bibfield  {author} {\bibinfo {author} {\bibfnamefont {E.}~\bibnamefont
  {Ridolfi}}, \bibinfo {author} {\bibfnamefont {D.}~\bibnamefont {Le}},
  \bibinfo {author} {\bibfnamefont {T.}~\bibnamefont {Rahman}}, \bibinfo
  {author} {\bibfnamefont {E.}~\bibnamefont {Mucciolo}}, \ and\ \bibinfo
  {author} {\bibfnamefont {C.}~\bibnamefont {Lewenkopf}},\ }\href@noop {}
  {\bibfield  {journal} {\bibinfo  {journal} {Journal of Physics: Condensed
  Matter}\ }\textbf {\bibinfo {volume} {27}},\ \bibinfo {pages} {365501}
  (\bibinfo {year} {2015})}\BibitemShut {NoStop}%
\bibitem [{\citenamefont {Ridolfi}\ \emph {et~al.}(2018)\citenamefont
  {Ridolfi}, \citenamefont {Lewenkopf},\ and\ \citenamefont
  {Pereira}}]{ridolfi2018exstruc}%
  \BibitemOpen
  \bibfield  {author} {\bibinfo {author} {\bibfnamefont {E.}~\bibnamefont
  {Ridolfi}}, \bibinfo {author} {\bibfnamefont {C.~H.}\ \bibnamefont
  {Lewenkopf}}, \ and\ \bibinfo {author} {\bibfnamefont {V.~M.}\ \bibnamefont
  {Pereira}},\ }\href {\doibase 10.1103/PhysRevB.97.205409} {\bibfield
  {journal} {\bibinfo  {journal} {Phys. Rev. B}\ }\textbf {\bibinfo {volume}
  {97}},\ \bibinfo {pages} {205409} (\bibinfo {year} {2018})}\BibitemShut
  {NoStop}%
\bibitem [{\citenamefont {Berkelbach}\ \emph {et~al.}(2013)\citenamefont
  {Berkelbach}, \citenamefont {Hybertsen},\ and\ \citenamefont
  {Reichman}}]{PhysRevB.88.045318}%
  \BibitemOpen
  \bibfield  {author} {\bibinfo {author} {\bibfnamefont {T.~C.}\ \bibnamefont
  {Berkelbach}}, \bibinfo {author} {\bibfnamefont {M.~S.}\ \bibnamefont
  {Hybertsen}}, \ and\ \bibinfo {author} {\bibfnamefont {D.~R.}\ \bibnamefont
  {Reichman}},\ }\href {\doibase 10.1103/PhysRevB.88.045318} {\bibfield
  {journal} {\bibinfo  {journal} {Phys. Rev. B}\ }\textbf {\bibinfo {volume}
  {88}},\ \bibinfo {pages} {045318} (\bibinfo {year} {2013})}\BibitemShut
  {NoStop}%
\bibitem [{\citenamefont {Vidal}(2003)}]{vidal2003efficient}%
  \BibitemOpen
  \bibfield  {author} {\bibinfo {author} {\bibfnamefont {G.}~\bibnamefont
  {Vidal}},\ }\href@noop {} {\bibfield  {journal} {\bibinfo  {journal} {Phys.
  Rev. Lett.}\ }\textbf {\bibinfo {volume} {91}},\ \bibinfo {pages} {147902}
  (\bibinfo {year} {2003})}\BibitemShut {NoStop}%
\bibitem [{\citenamefont {Or{\'u}s}(2014)}]{orus2014practical}%
  \BibitemOpen
  \bibfield  {author} {\bibinfo {author} {\bibfnamefont {R.}~\bibnamefont
  {Or{\'u}s}},\ }\href@noop {} {\bibfield  {journal} {\bibinfo  {journal}
  {Annals of Physics}\ }\textbf {\bibinfo {volume} {349}},\ \bibinfo {pages}
  {117} (\bibinfo {year} {2014})}\BibitemShut {NoStop}%
\bibitem [{\citenamefont {Schollw{\"o}ck}(2011)}]{schollwock2011density}%
  \BibitemOpen
  \bibfield  {author} {\bibinfo {author} {\bibfnamefont {U.}~\bibnamefont
  {Schollw{\"o}ck}},\ }\href@noop {} {\bibfield  {journal} {\bibinfo  {journal}
  {Annals of Physics}\ }\textbf {\bibinfo {volume} {326}},\ \bibinfo {pages}
  {96} (\bibinfo {year} {2011})}\BibitemShut {NoStop}%
\bibitem [{\citenamefont {Vidal}(2004)}]{vidal2004efficient}%
  \BibitemOpen
  \bibfield  {author} {\bibinfo {author} {\bibfnamefont {G.}~\bibnamefont
  {Vidal}},\ }\href@noop {} {\bibfield  {journal} {\bibinfo  {journal} {Phy.
  Rev. Lett.}\ }\textbf {\bibinfo {volume} {93}},\ \bibinfo {pages} {040502}
  (\bibinfo {year} {2004})}\BibitemShut {NoStop}%
\bibitem [{\citenamefont {Tietze}\ \emph {et~al.}(2015)\citenamefont {Tietze},
  \citenamefont {Schenk},\ and\ \citenamefont {Gamm}}]{tietze2015electronic}%
  \BibitemOpen
  \bibfield  {author} {\bibinfo {author} {\bibfnamefont {U.}~\bibnamefont
  {Tietze}}, \bibinfo {author} {\bibfnamefont {C.}~\bibnamefont {Schenk}}, \
  and\ \bibinfo {author} {\bibfnamefont {E.}~\bibnamefont {Gamm}},\ }\href@noop
  {} {\emph {\bibinfo {title} {Electronic circuits: handbook for design and
  application}}}\ (\bibinfo  {publisher} {Springer},\ \bibinfo {year}
  {2015})\BibitemShut {NoStop}%
\bibitem [{\citenamefont {Stoudenmire}\ and\ \citenamefont
  {White}()}]{itensor}%
  \BibitemOpen
  \bibfield  {author} {\bibinfo {author} {\bibfnamefont {E.~M.}\ \bibnamefont
  {Stoudenmire}}\ and\ \bibinfo {author} {\bibfnamefont {S.~R.}\ \bibnamefont
  {White}},\ }\href {http://itensor.org/} {\enquote {\bibinfo {title} {Itensor
  c++ library,http://itensor.org/},}\ }\BibitemShut {NoStop}%
\bibitem [{\citenamefont {White}(1992)}]{white1992density}%
  \BibitemOpen
  \bibfield  {author} {\bibinfo {author} {\bibfnamefont {S.~R.}\ \bibnamefont
  {White}},\ }\href@noop {} {\bibfield  {journal} {\bibinfo  {journal}
  {Physical review letters}\ }\textbf {\bibinfo {volume} {69}},\ \bibinfo
  {pages} {2863} (\bibinfo {year} {1992})}\BibitemShut {NoStop}%
\bibitem [{\citenamefont {Qiu}\ \emph {et~al.}(2016)\citenamefont {Qiu},
  \citenamefont {da~Jornada},\ and\ \citenamefont
  {Louie}}]{PhysRevB.93.235435}%
  \BibitemOpen
  \bibfield  {author} {\bibinfo {author} {\bibfnamefont {D.~Y.}\ \bibnamefont
  {Qiu}}, \bibinfo {author} {\bibfnamefont {F.~H.}\ \bibnamefont {da~Jornada}},
  \ and\ \bibinfo {author} {\bibfnamefont {S.~G.}\ \bibnamefont {Louie}},\
  }\href {\doibase 10.1103/PhysRevB.93.235435} {\bibfield  {journal} {\bibinfo
  {journal} {Phys. Rev. B}\ }\textbf {\bibinfo {volume} {93}},\ \bibinfo
  {pages} {235435} (\bibinfo {year} {2016})}\BibitemShut {NoStop}%
\bibitem [{\citenamefont {Lim}\ and\ \citenamefont
  {Sheng}(2016)}]{PhysRevB.94.045111}%
  \BibitemOpen
  \bibfield  {author} {\bibinfo {author} {\bibfnamefont {S.~P.}\ \bibnamefont
  {Lim}}\ and\ \bibinfo {author} {\bibfnamefont {D.~N.}\ \bibnamefont
  {Sheng}},\ }\href {\doibase 10.1103/PhysRevB.94.045111} {\bibfield  {journal}
  {\bibinfo  {journal} {Phys. Rev. B}\ }\textbf {\bibinfo {volume} {94}},\
  \bibinfo {pages} {045111} (\bibinfo {year} {2016})}\BibitemShut {NoStop}%
\bibitem [{\citenamefont {Yu}\ \emph {et~al.}(2017)\citenamefont {Yu},
  \citenamefont {Pekker},\ and\ \citenamefont
  {Clark}}]{PhysRevLett.118.017201}%
  \BibitemOpen
  \bibfield  {author} {\bibinfo {author} {\bibfnamefont {X.}~\bibnamefont
  {Yu}}, \bibinfo {author} {\bibfnamefont {D.}~\bibnamefont {Pekker}}, \ and\
  \bibinfo {author} {\bibfnamefont {B.~K.}\ \bibnamefont {Clark}},\ }\href
  {\doibase 10.1103/PhysRevLett.118.017201} {\bibfield  {journal} {\bibinfo
  {journal} {Phys. Rev. Lett.}\ }\textbf {\bibinfo {volume} {118}},\ \bibinfo
  {pages} {017201} (\bibinfo {year} {2017})}\BibitemShut {NoStop}%
\bibitem [{\citenamefont {Khemani}\ \emph {et~al.}(2016)\citenamefont
  {Khemani}, \citenamefont {Pollmann},\ and\ \citenamefont
  {Sondhi}}]{PhysRevLett.116.247204}%
  \BibitemOpen
  \bibfield  {author} {\bibinfo {author} {\bibfnamefont {V.}~\bibnamefont
  {Khemani}}, \bibinfo {author} {\bibfnamefont {F.}~\bibnamefont {Pollmann}}, \
  and\ \bibinfo {author} {\bibfnamefont {S.~L.}\ \bibnamefont {Sondhi}},\
  }\href {\doibase 10.1103/PhysRevLett.116.247204} {\bibfield  {journal}
  {\bibinfo  {journal} {Phys. Rev. Lett.}\ }\textbf {\bibinfo {volume} {116}},\
  \bibinfo {pages} {247204} (\bibinfo {year} {2016})}\BibitemShut {NoStop}%
\end{thebibliography}

%merlin.mbs apsrev4-1.bst 2010-07-25 4.21a (PWD, AO, DPC) hacked
%Control: key (0)
%Control: author (8) initials jnrlst
%Control: editor formatted (1) identically to author
%Control: production of article title (-1) disabled
%Control: page (0) single
%Control: year (1) truncated
%Control: production of eprint (0) enabled
%

%TC:endignore

\end{document}